\documentclass[journal]{./IEEEtran}

\ifCLASSINFOpdf
  \usepackage[pdftex]{graphicx}
\else
 
\fi

\usepackage{amsmath}
\usepackage{amssymb}
\usepackage{amsfonts}

\usepackage{subfig}

\usepackage{multicol}

\begin{document}

\newtheorem{theorem}{Theorem}
\newtheorem{definition}{Definition}

\title{Combining Adaptive Coding and Modulation with Hierarchical Modulation in Satcom Systems}

\author{\IEEEauthorblockN{Hugo M{\'e}ric, 
J{\'e}r{\^o}me Lacan, 
Fabrice Arnal, 
Guy Lesthievent and 
Marie-Laure Boucheret
\IEEEcompsocitemizethanks{H. M{\'e}ric and J. Lacan are with Universit{\'e} de Toulouse/ISAE and T{\'e}SA, Toulouse, France (e-mail: hugo.meric@isae.fr; jerome.lacan@isae.fr).\protect\\
F. Arnal is with Thales Alenia Space, Toulouse, France (e-mail: fabrice.arnal@thalesaleniaspace.com).\protect\\
G. Lesthievent is with CNES, Toulouse, France (e-mail: guy.lesthievent@cnes.fr).\protect\\
M.-L. Boucheret is with Universit{\'e} de Toulouse/ENSEEIHT and T{\'e}SA, Toulouse, France (e-mail: marie-laure.boucheret@enseeiht.fr).
}}}


\maketitle

\begin{abstract}
We investigate the design of a broadcast system in order to maximize throughput. This task is usually challenging due to channel variability. Forty years ago, Cover introduced and compared two schemes: time sharing and superposition coding. Even if the second scheme was proved to be optimal for some channels, modern satellite communications systems such as DVB-SH and DVB-S2 rely mainly on a time sharing strategy to optimize the throughput. They consider hierarchical modulation, a practical implementation of superposition coding, but only for unequal error protection or backward compatibility purposes. In this article, we propose to combine time sharing and hierarchical modulation together and show how this scheme can improve the performance in terms of available rate. We introduce a hierarchical 16-APSK to boost the performance of the DVB-S2 standard. We also evaluate various strategies to group the receivers in pairs when using hierarchical modulation. Finally, we show in a realistic case, based on DVB-S2, that the combined scheme can provide throughput gains greater than 10\% compared to the best time sharing strategy.
\end{abstract}


\IEEEpeerreviewmaketitle

\section{Introduction}

In most broadcast applications, the Signal-to-Noise Ratio (SNR) experienced by each receiver can be quite different. For instance, in satellite communications the channel quality decreases with the presence of clouds in Ku or Ka band, or with shadowing effects of the environment in lower bands. The first solution for broadcasting was to design the system for the worst-case reception, but this leads to poor performance as many receivers do not exploit their full potential. Two other schemes were proposed in \cite{cover} and \cite{bergmans}: time division multiplexing with variable coding and modulation, and superposition coding. Time division multiplexing, or time sharing, allocates a proportion of time to communicating with each receiver using any modulation and error protection level. This functionality, called Variable Coding and Modulation (VCM) \cite{s2}, is in practice the most used in standards today. If a return channel is available, VCM may be combined with Adaptive Coding and Modulation (ACM) to optimize the transmission parameters \cite{s2}. In superposition coding, the available energy is shared among several service flows which are sent simultaneously in the same band.
\begin{figure}[!t]
\centering
\includegraphics[width = 0.95\columnwidth]{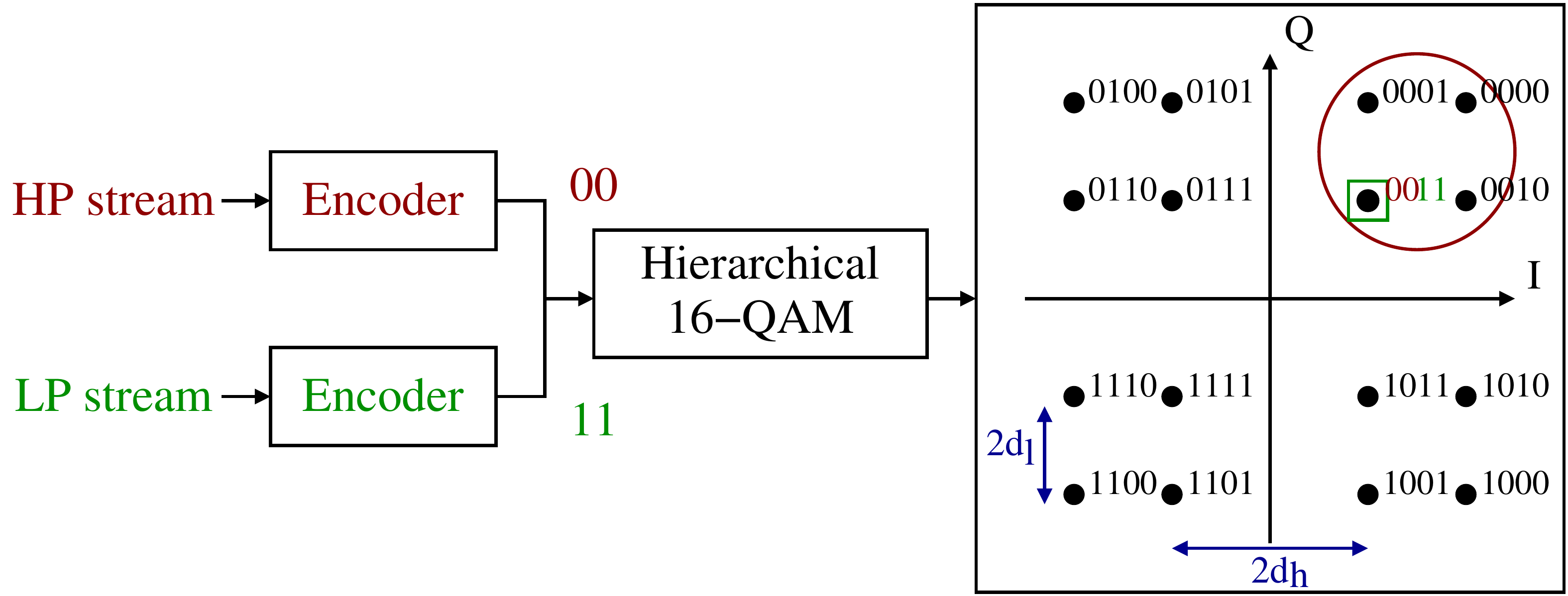}
\caption{Hierarchical modulation using a 16-QAM}
\label{hm_principle}
\end{figure}
This scheme was introduced by Cover in \cite{cover} in order to improve the transmission rate from a single source to several receivers. When communicating with two receivers, the principle is to superimpose information for the user with the best SNR. This superposition can be done directly at the Forward Error Correction (FEC) level or at the modulation level as shown in \figurename~\ref{hm_principle} with a 16 Quadrature Amplitude Modulation (16-QAM).

Hierarchical modulation is a practical implementation of superposition coding. Although hierarchical modulation has been introduced to improve throughput, it is currently often used to provide unequal protection. The idea is to merge two different streams at the modulation level. The High Priority (HP) stream is used to select the quadrant, and the Low Priority (LP) stream selects the position inside the quadrant. The HP stream is dedicated to users with poor channel quality, unlike the LP stream which requires a large SNR to be decoded error-free. In \cite{svc_hm}, video encoded with the scalable video coding extension of the H.264 standard \cite{svc} is protected using hierarchical modulation. The base layer of the video is transmitted in the HP stream, while the enhanced layer is carried by the LP stream. Another usage is backward compatibility \cite{backward_compatibility}, \cite{broad05}. The DVB-S2\footnote{Digital Video Broadcasting - Satellite - Second Generation} standard \cite{s2}, \cite{dvbs2} is called upon to replace the DVB-S\footnote{Digital Video Broadcasting - Satellite} standard, but many DVB-S receivers are already installed. Thus, the hierarchical modulation helps the migration by allowing the DVB-S receivers to operate. In \cite{local_content}, the authors propose to provide local content with hierarchical modulation. The principle is to carry local information that is of interest to a particular geographic area in the LP stream, while the HP stream transmits global content. Other works improve the performance of relay communication system \cite{relaycom} or OFDMA-based networks \cite{icc10}. Finally, multilevel codes are another way to deal with broadcast channels using hierarchical transmission \cite{schill,vargas}.

Our work focuses on using hierarchical modulation in modern broadcast systems to increase the transmission rate. For instance, even if the Low-Density Parity-Check (LDPC) codes of DVB-S2 approach the Shannon limit for the Additive White Gaussian Noise (AWGN) channel with one receiver \cite{SAT:SAT787}, the throughput can be greatly increased for the broadcast case. Indeed, Cover presents the set of achievable rates for the Gaussian broadcast channel with two receivers in \cite{cover}. This set clearly dominates the time sharing achievable rate. Our article investigates the performance, in terms of throughput, of a satellite broadcast system where time sharing and hierarchical modulation are combined. We show in an example modeling a satellite broadcast area that the gain can be significant. A way of grouping receivers in pairs is also investigated as it greatly affects performance.
 
The paper is organized as follows: Section~\ref{part2} presents the hierarchical modulation. We introduce the hierarchical 16 Amplitude and
Phase-Shift Keying (16-APSK) modulation in order to boost the performance of the DVB-S2 standard. In Section~\ref{part3}, the achievable rates for an AWGN channel are computed for the time sharing used alone or combined with hierarchical modulation. We study on a use case the performance of each scheme in Section~\ref{part4}. We also propose a grouping strategy when using hierarchical modulation and discuss its performance. Finally, Section~\ref{part5} concludes the paper by summarizing the results.

\section{Hierarchical Modulation}\label{part2}

This part introduces the hierarchical 16-QAM and 16-APSK. First, the hierarchical 16-QAM, considered in the DVB-SH\footnote{Digital Video Broadcasting - Satellite to Handheld} standard \cite{sh},\cite{DVBSH}, is presented. This gives us some insight on how to introduce the hierarchical 16-APSK which is presented in order to improve the performance of DVB-S2. As mentioned before, hierarchical modulations merge several streams in a same symbol. The available energy is shared between each stream. In our study, two streams are considered. When hierarchical modulation is used for unequal protection purposes, these flows are called HP and LP streams. However, unequal protection is not the goal of our work, so we will now refer to High Energy (HE) and Low Energy (LE) streams for the streams containing the most and the least energy, respectively.

As each stream does not use the same energy, hierarchical modulations are based on non-uniform constellations where the symbols are not uniformly distributed in the space. The geometry of non-uniform constellations is described using the constellation parameter(s).

\subsection{Hierarchical 16-QAM}
The constellation parameter $\alpha$ is defined by $d_h/d_l$, where $2d_h$ is the minimum distance between two constellation points carrying different HE bits and $2d_l$ is the minimum distance between any constellation point (see \figurename~\ref{hm_principle}). Typically, we have $\alpha \geqslant 1$, where $\alpha=1$ corresponds to the uniform 16-QAM, but it is also possible to have $\alpha \leqslant 1$ \cite{icssc11}. At a given energy per symbol ($E_s$), when $\alpha$ grows, the constellation points in each quadrant become farther from the I and Q axes. Thus it is easier to decode the HE stream. However, in the same quadrant, the points become closer and the LE stream requires a better channel quality to be decoded error-free.

For Quadrature Phase-Shift Keying (QPSK) modulation, we define the constellation parameter as the minimum distance between two constellation points. The hierarchical 16-QAM is the superposition of two QPSK modulations, one with parameter $2(d_h+d_l)$ (carrying the HE stream) and the other with parameter $2d_l$ (carrying the LE stream). The energy ratio between the two streams is
\begin{eqnarray}
\frac{E_{he}}{E_{le}} = (1+\alpha)^2 ,
\end{eqnarray}

\noindent where $E_{he}$ and $E_{le}$ correspond to the amount of energy allocated to the HE and LE streams, respectively \cite{local_content}. The DVB-SH standard recommends two values for $\alpha$: 2 and 4. In fact, it also considers $\alpha=1$ but only in the VCM mode. The values 2 and 4 are defined in order to provide unequal protection. From an energy point of view, this amounts to giving 90\% ($\alpha=2$) or 96\% ($\alpha=4$) of the available energy to the HE stream. In \cite{icc12}, the authors improve the overall throughput of a simple broadcast channel by adding $\alpha=1$, $\alpha=0.8$ and $\alpha=0.5$. This provides a better repartition of energy: the HE stream then contains 80\%, 76\% and 69\% of the total power, respectively. Note that when $E_{he}=E_{le}$ (i.e., each stream contains 50\% of the total power), it is equivalent to superposing two QPSK modulations with the same energy and the resulting hierarchical 16-QAM has a constellation parameter of $\alpha=0$.

\subsection{Hierarchical 16-APSK}
\begin{figure}[!ht]
\centering
\includegraphics[width = 0.5\columnwidth]{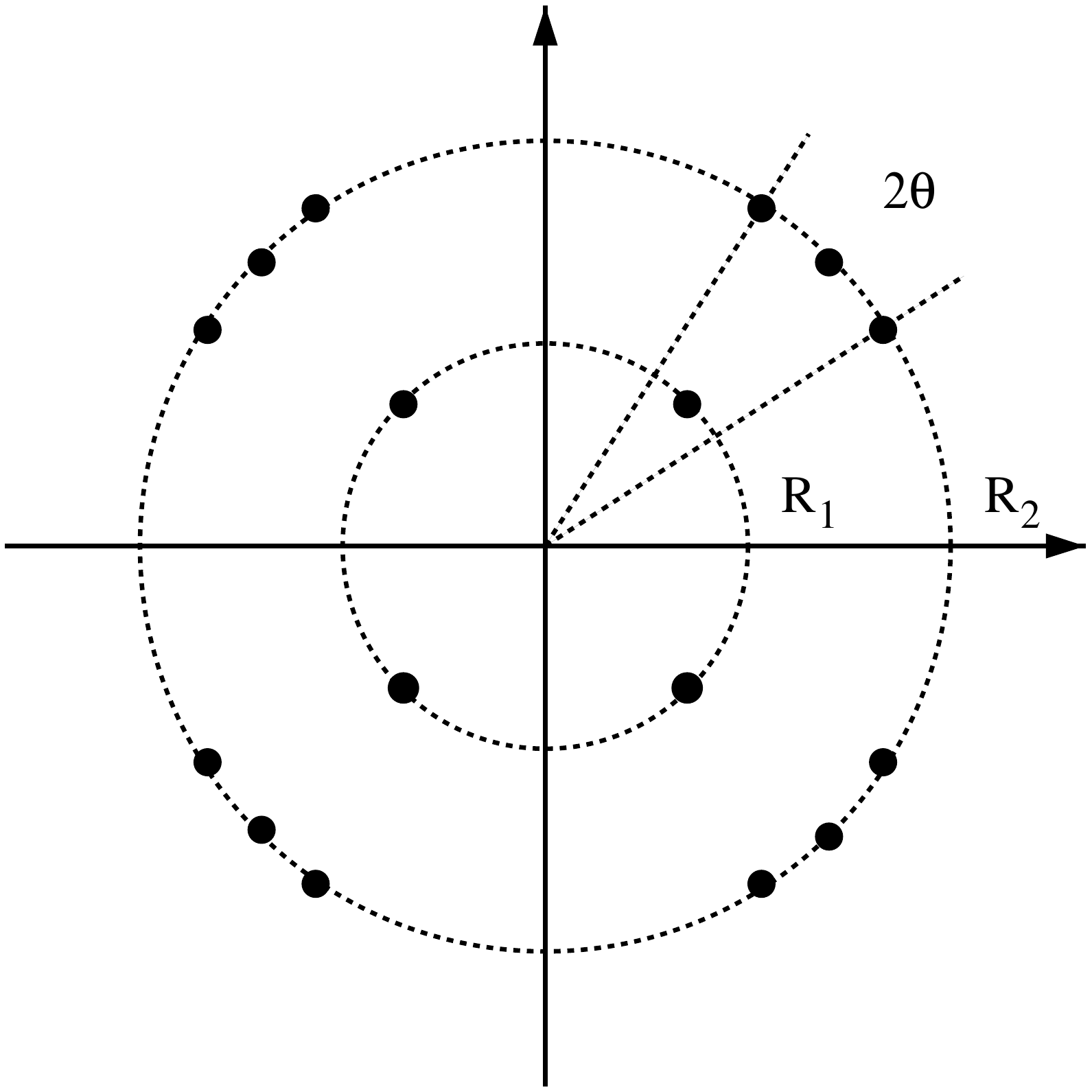}
\caption{16-APSK modulation}
\label{16apskh}
\end{figure}

The DVB-S2 standard also introduces hierarchical modulation with the hierarchical 8-PSK. The constellation parameter $\theta_{PSK}$, which is the half angle between two points in one quadrant, is defined by the service operator according to the desired performance. However, this modulation does not offer a good spectrum efficiency. As the 16-APSK is already defined in DVB-S2, we propose the hierarchical 16-APSK, shown in \figurename~\ref{16apskh}, in order to boost system performance. The constellation parameters are the ratio between the radius of the outer ($R_2$) and inner ($R_1$) rings $\gamma = R_2/R_1$ , and the half angle between the points on the outer ring in each quadrant $\theta$ (see \figurename~\ref{16apskh}).

The hierarchical 16-APSK is not a new concept. For instance, this modulation is presented to upgrade an existing digital broadcast system in \cite{channel_capacity_enhancement}. However, the design of the modulation is not addressed. In \cite{SAT:SAT841}, the authors investigate the design of APSK modulations for satellite broadband communication, but the hierarchical case is not treated. In our study, we use an energy argument to choose the parameters $\gamma$ and $\theta$. The hierarchical modulation shares the available energy between the HE and LE streams. We consider the energy of the HE stream, given by the energy of a QPSK modulation where the constellation points are located at the barycenter of the four points in each quadrant. Using the polar coordinates, the barycenter in the upper right quadrant is
\begin{equation}
z_b = e^{i\pi/4}\frac{R_1+R_2+2R_2\cos(\theta)}{4}.
\label{barycentre}
\end{equation}

Moreover, the symbol energy is expressed as
\begin{equation}
E_s = \frac{4R_1^2+12R_2^2}{16} = \frac{1+3\gamma^2}{4}R_1^2.
\label{Es}
\end{equation}

Then combining (\ref{barycentre}) and (\ref{Es}), the distance of the barycenter to the origin is
\begin{equation}
d_B = \left| z_b \right| = \frac{1+\gamma(1+2\cos(\theta))}{4} \frac{2\sqrt{E_s}}{\sqrt{1+3\gamma^2}}.
\label{module}
\end{equation}

Finally, the energy of the HE stream is  given by
\begin{eqnarray}
E_{he} & = & E_{qpsk} \nonumber \\
		& = & d_B^2 \nonumber \\
		& = & \underbrace{ \frac{\left( 1+\gamma (1+2\cos\theta) \right)^2}{4(1+3\gamma^2)} }_{\rho_{he}} E_s.
\label{energie}
\end{eqnarray}

Equation (\ref{energie}) introduces $\rho_{he}$ as the ratio between the energy of the HE stream $E_{he}$ and the symbol energy $E_s$. As the HE stream contains more energy than the LE stream, we verify that $\rho_{he} \geqslant 0.5$. We are now interested in determining the set of $(\gamma, \theta)$ pairs which are solutions of  
\begin{equation}
\rho_{he}  = \frac{\left( 1+\gamma (1+2\cos\theta) \right)^2}{4(1+3\gamma^2)},
\label{eq_energie}
\end{equation}
where $\rho_{he} \geqslant 0.5$ is known, $\gamma \geqslant 1$ and $\theta \geqslant 0$. The resolution of (\ref{eq_energie}) is given in Appendix~\ref{resolution_equation}. The solution set is
\begin{equation}
\mathcal{S}_{\rho_{he}} = \left\{ \left( \gamma, \arccos \left( f(\gamma, \rho_{he}) \right) \right) | 1 \leqslant \gamma \leqslant \gamma_{lim} \right\},
\label{sol}
\end{equation}
where
\begin{equation}
f(\gamma, \rho_{he}) = \frac{1}{2} \left( \frac{\sqrt{4\rho_{he}(1+3\gamma^2)}-1}{\gamma} -1 \right)
\end{equation}
and
\begin{equation}
\gamma_{lim} = \begin{cases}  +\infty, & \text{if } \rho_{he}\leqslant 0.75 \\ \frac{3+4\sqrt{3\rho_{he}(1-\rho_{he})}}{3(4\rho_{he}-3)}, & \text{if } \rho_{he} > 0.75.  \end{cases}
\end{equation}

Appendix~\ref{resolution_equation} also presents two examples of the $\mathcal{S}_{\rho_{he}}$ set. When $\rho_{he}$ increases, the points in one quadrant tend to become closer. This implies that the HE stream is easier to decode, whereas the LE stream requires a better SNR to be decoded error-free. As for the hierarchical 16-QAM, several values of $\rho_{he}$ have to be selected. The results presented in Section~\ref{part4} show that choosing $\rho_{he}=0.75, 0.8, 0.85 \text{ and } 0.9$ allows us to achieve a maximum rate gain of roughly 15\%. However, once the parameter $\rho_{he}$ is set, we still have to decide which $(\gamma, \theta)$ pair to keep in the $\mathcal{S}_{\rho_{he}}$ set. We keep one $(\gamma, \theta)$ pair per $\rho_{he}$ value. This pair minimizes the average decoding threshold for the HE stream over all the coding rates in the DVB-S2 standard. Appendix~\ref{appendixB} gives all the decoding thresholds and explains their computations.

\section{Time Sharing and Hierarchical Modulation Achievable Rates}\label{part3}

This part introduces the achievable rates over an AWGN channel by the two following schemes: time sharing with or without hierarchical modulation, referred to as \emph{hierarchical modulation} and \emph{classical} time sharing, respectively. We first consider the case of one source communicating with two receivers, then we study the general case for $n$ receivers. In both cases, we assume that the transmitter has knowledge of the SNR at the receivers.

\subsection{Achievable rates: case with two receivers}

\subsubsection{Classical time sharing}
We consider one source communicating with two receivers, each one with a particular SNR. Given this SNR, we assume that receiver $i$ ($i=1,2$) has a rate $R_i$, which corresponds to the best rate it can manage. This rate is the amount of useful data transmitted on the link. It is dependent on the modulations and coding rates available in the system. For instance, if the modulation is a QPSK and the code rate is 1/3, then the rate equals $2 \times 1/3$ bit/symbol. In our study, the physical layer is based on the DVB-S2 standard \cite{dvbs2}.

The time sharing scheme allocates a fraction of time $t_i$ to receiver $i$. We define the average rate for receiver $i$ as $t_iR_i$. In our study, we are interested in offering the \emph{same average rate} to everyone, but our work can be easily adapted to another rate policy. To offer the same rate to both users, we need to solve
\begin{equation}
\begin{cases} t_1R_1 = t_2R_2 \\ t_1 + t_2 = 1. \end{cases}
\label{sh_equation_1}
\end{equation} 

By solving (\ref{sh_equation_1}), the fraction of time allocated to each user is
\begin{equation}
\begin{cases} t_1 = \frac{R_2}{R_1+R_2} \\ t_2 = \frac{R_1}{R_1+R_2}. \end{cases}
\label{fraction_time_1}
\end{equation}

The constraint $t_1 + t_2 = 1$ is verified and we remark that increasing $R_i$ reduces $t_i$, which is a consequence of our rate policy. Finally, the rate offered to each receiver is
\begin{equation}
R_{ts} = \frac{R_1R_2}{R_1+R_2}.
\label{rate_1}
\end{equation}

\subsubsection{Hierarchical modulation time sharing}
The first step is to compute the rates offered by all the possible modulations, including hierarchical ones. When hierarchical modulation is used, we assume that \emph{the receiver experiencing the best SNR decodes the LE stream}. Thus, we obtain a set of operating points. When two sets of rates $(R_1,R_2)$ and $\left( R_1^*,R_2^* \right)$ are available, the time sharing strategy allows any rate pair 
\begin{equation}
\left(\tau R_1+(1-\tau)R_1^* , \tau R_2+(1-\tau)R_2^* \right),
\end{equation}
where $0 \leqslant \tau \leqslant 1$ is the fraction of time allocated to $(R_1,R_2)$. The achievable rates set finally corresponds to the convex hull of all the operating points. As we are interested in offering the \emph{same rate} to the users, we calculate the intersection of the convex hull with the curve $y=x$. We note $R_{hm}$ and $R_{ts}$ as the rates offered to both receivers by the hierarchical modulation and classical time sharing strategy, respectively. 

\begin{figure}[!ht]
\centering
\includegraphics[width = 0.9\columnwidth]{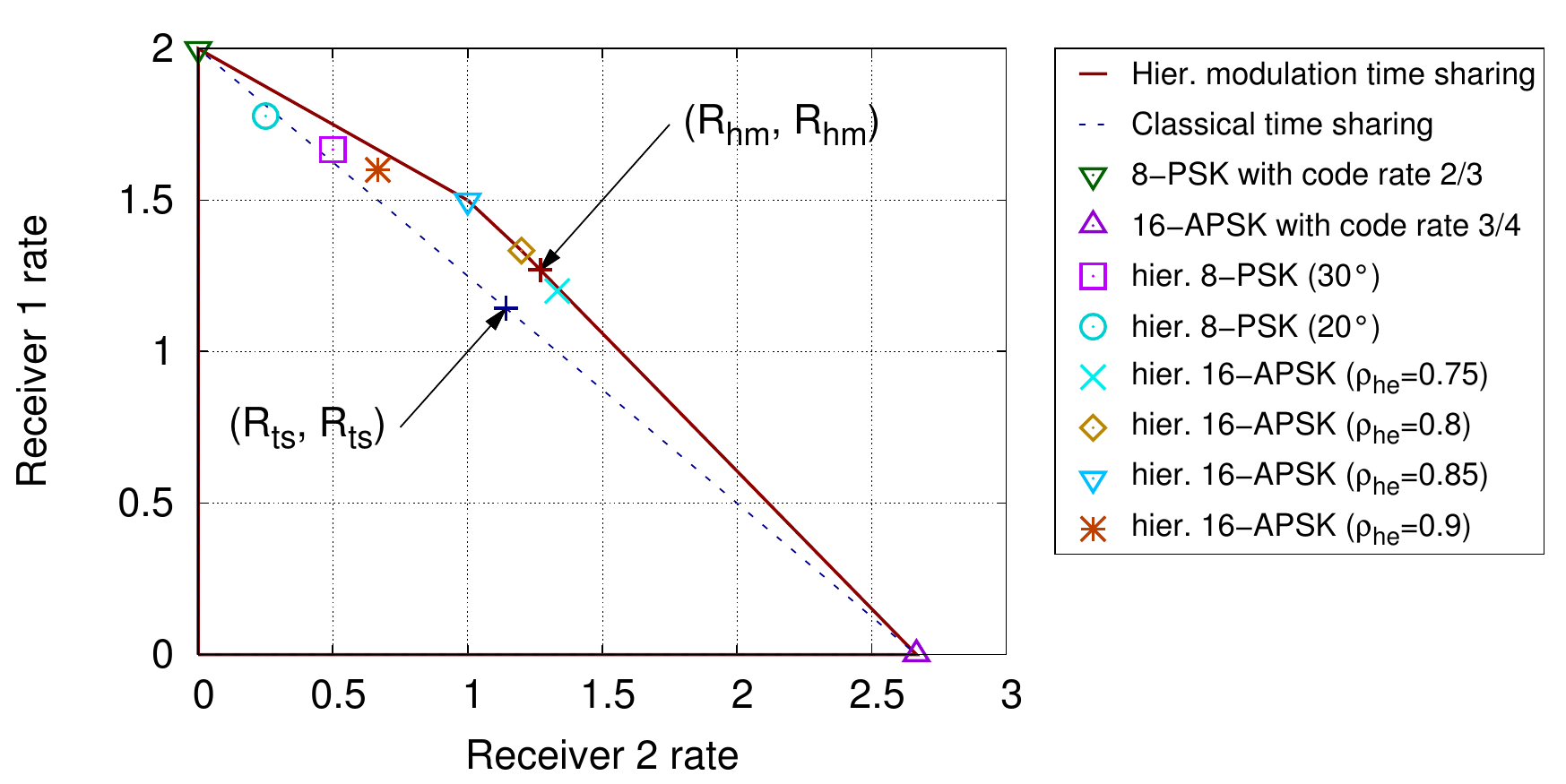}
\caption{Achievable rates set: $\text{SNR}_1=7\text{ dB}$ - $\text{SNR}_2=10\text{ dB}$}
\label{achievable_rates}
\end{figure}

\figurename~\ref{achievable_rates} presents one example of an achievable rates set when one receiver experiences a SNR of 7 dB and the other 10 dB. We also represent $R_{hm}$ and $R_{ts}$ in order to visualize the gain. For the classical time sharing strategy, the rates obtained in \figurename~\ref{achievable_rates} result from the transmission with the LDPC code of rate 2/3 associated with 8-PSK modulation for the user with a SNR of 7 dB and the LDPC code of rate 3/4 associated with 16-APSK modulation for the other user. As said before, it is the most that each receiver can manage. For the hierarchical modulation, the operating points are computed using the standard \cite{dvbs2} or as in Appendix~\ref{appendixB}. Remark that the hierarchical 16-APSK gives better results than the hierarchical 8-PSK. Moreover, when $\rho_{he}$ increases (for a given code rate), more energy is dedicated to the HE stream and it can be decoded error-free with a smaller SNR. Thus, the user with the worst SNR has a rate which increases. However, less energy is allocated in the LE stream and its performance decreases. Finally, the interest of using hierarchical modulation is evident as the gain between $R_{hm}$ and $R_{ts}$ is about 11\%. 

\begin{figure}[!ht]
\centering
\includegraphics[width = 0.9\columnwidth]{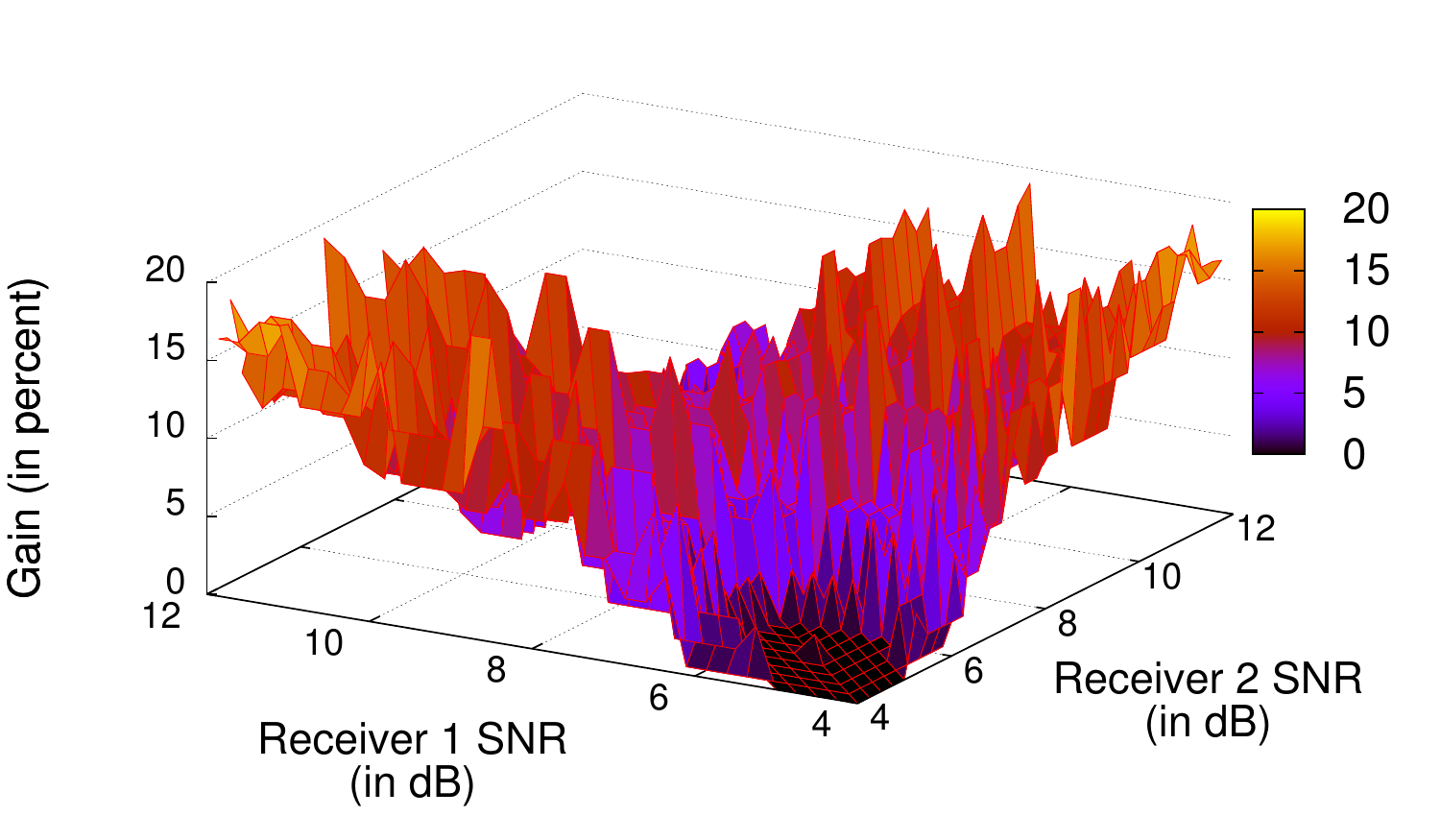}
\caption{Rate gain}
\label{rate_gain}
\end{figure}

We now use the same method to evaluate the gain between $R_{hm}$ and $R_{ts}$ for all $(\text{SNR}_1, \text{SNR}_2)$ pairs where $4\text{ dB} \leqslant \text{SNR}_i \leqslant 12\text{ dB}$ and $i=1,2$. \figurename~\ref{rate_gain} presents the results. Note that the gain provided by hierarchical modulation is significant in several cases and can achieve up to 20\% of rate improvement. In general, the gain becomes greater as the SNR difference between the two receivers increases. This observation will be used in the next part when we group a set of users in pairs.

\subsection{Achievable rates: case with $n$ receivers}

\subsubsection{Classical time sharing}

We now consider a broadcast system with $n$ receivers. We assume that receiver $i$ has a rate $R_i$, which corresponds to the best rate it can manage as mentioned above. With our rate policy, (\ref{sh_equation_1}) becomes
\begin{equation}
\begin{cases} \forall i,j, \text{ } t_iR_i = t_jR_j \\ \sum_{i} t_{i} = 1. \end{cases}
\label{sh_equation}
\end{equation} 

The resolution of (\ref{sh_equation}) leads to a fraction of time allocated to user $i$ of
\begin{equation}
t_{i} = \frac{\prod_{k\neq i} R_k}{\sum_{j=1}^n \left(\prod_{k\neq j} R_k \right)}.
\label{fraction_time}
\end{equation}

With this time allocation, the average rate offered to each receiver is
\begin{equation}
R_{ts} = \frac{\prod_{k} R_k}{\sum_{j=1}^n \left(\prod_{k\neq j} R_k \right)} = \left( \sum_{j=1}^n \frac{1}{R_j}\right)^{-1}.
\label{rate}
\end{equation}

\subsubsection{Hierarchical modulation with time sharing}
For the case of $n$ receivers, the first step is to group the users in pairs in order to use hierarchical modulation. A lot of possibilities are available and the next section presents a grouping strategy which generally obtains good results. Once the pairs have been chosen, we compute for each pair the achievable rate as previously described. Finally, we need to equalize the rate between each user. This is done using time sharing (\ref{fraction_time}). For instance, consider a DVB-S2 system where a user $u_1$ with a SNR of 7 dB is paired with a user $u_2$ with a SNR of 10 dB. The rate for each receiver is obtained using the hierarchical 16-APSK ($\rho_{he}=0.8$) a fraction of time $a_1$ and the 16-APSK (considered in the DVB-S2 standard) a fraction of time $a_2$ as shown in \figurename~\ref{achievable_rates}. When equalizing the rates between all the users, (\ref{fraction_time}) gives the same fraction of time $t$ to users $u_1$ and $u_2$. Then, the pair of users has to share a global fraction of time $2t$. It follows that the broadcast system allocates to $u_1$ and $u_2$ the hierarchical 16-APSK ($\rho_{he}=0.8$) for a time proportion $2t \times a_1$ and the 16-APSK for $2t \times a_2$.

\section{Application to broadcast channel}\label{part4}

An important consequence of the previous section is that the gain from hierarchical modulation depends on how users are paired. For instance, consider a set of four users $u_1$, $u_2$, $u_3$, $u_4$ with respective SNRs 4, 4, 12 and 12 dB. Then, according to \figurename~\ref{rate_gain}, the choice of pairs $(u_1,u_2)$ and $(u_3,u_4)$ leads to no gain while the choice $(u_1,u_3)$ and $(u_2,u_4)$ provides a gain of about 20\%. In this section, we first present different grouping strategies for a set of users when hierarchical modulation time sharing is considered. Then we introduce an AWGN broadcast channel where the performance of hierarchical modulation and classical time sharing are evaluated. The impact of the grouping strategy is also discussed.

\subsection{Grouping strategy}

We consider a set of receivers, where the distribution of SNR values is known. The possible SNR values are $\text{SNR}_i$ with $1 \leqslant i \leqslant m$ and for all $i \leqslant j$, $\text{SNR}_i \leqslant \text{SNR}_j$. Moreover, exactly $n_i$ receivers experience a channel quality of $\text{SNR}_i$. We also define $\sum_{i=1}^m n_i=2N$ as the total number of receivers and $\Delta_{i,j}=|\text{SNR}_i-\text{SNR}_j|$.

\begin{definition}
For any  grouping, the average SNR difference per receivers in pairs is defined as
\begin{equation}
\Delta = \frac{1}{N} \sum_{k=1}^{N} \Delta_{i_k,j_k},
\label{average_delta}
\end{equation}
where the $(i_k,j_k)$ couple represents a pair of receivers.
\end{definition}

When communicating with two receivers, we have already mentioned that the gain is higher when the SNR difference between the two users is large. From this observation, we are looking to group the users in pairs in order to maximize the average SNR difference. The following theorem presents a strategy to compute this maximum. 

\begin{theorem}\label{our_theorem}
From any set of receivers, the iterative procedure that picks the two receivers with the largest SNR difference, grouping them and repeating this operation allows us to reach the maximum average SNR difference.
\end{theorem}

\begin{IEEEproof}
See Appendix~\ref{proof_theorem}.
\end{IEEEproof}

However, depending on the receivers' configuration, other schemes allow the maximum average SNR difference to be reached. For instance, consider the case of four receivers where the SNR values are $\text{SNR}_1$ (user 1), $\text{SNR}_2=\text{SNR}_1+1\text{ dB}$ (user 2), $\text{SNR}_3=\text{SNR}_1+2\text{ dB}$ (user 3) and $\text{SNR}_4=\text{SNR}_1+3\text{ dB}$ (user 4). The previous strategy leads to grouping user 1 with user 4, and user 2 with user 3. But it is also possible to group user 1 with user 3, and user 2 with user 4. In both cases, the average SNR difference is 2 dB. 

To highlight the impact of the grouping strategy, we propose comparing four grouping schemes:
\begin{itemize}
\item Strategy A: the scheme described in the previous theorem.
\item Strategy B: we compute the maximum average SNR difference $\Delta_{max}$ and use it to group the receivers with a SNR difference as close as possible to $\Delta_{max}$. This strategy usually allows an average SNR difference close to $\Delta_{max}$, but compared to strategy A, the variance of the SNR difference in pairs is much smaller.
\item Strategy C: the receivers are grouped randomly.
\item Strategy D: we group the receivers with the closest SNRs.
 \end{itemize}

\subsection{DVB-S2 channel model}

\begin{figure}[!ht]
\centering
\includegraphics[width = 0.45\columnwidth]{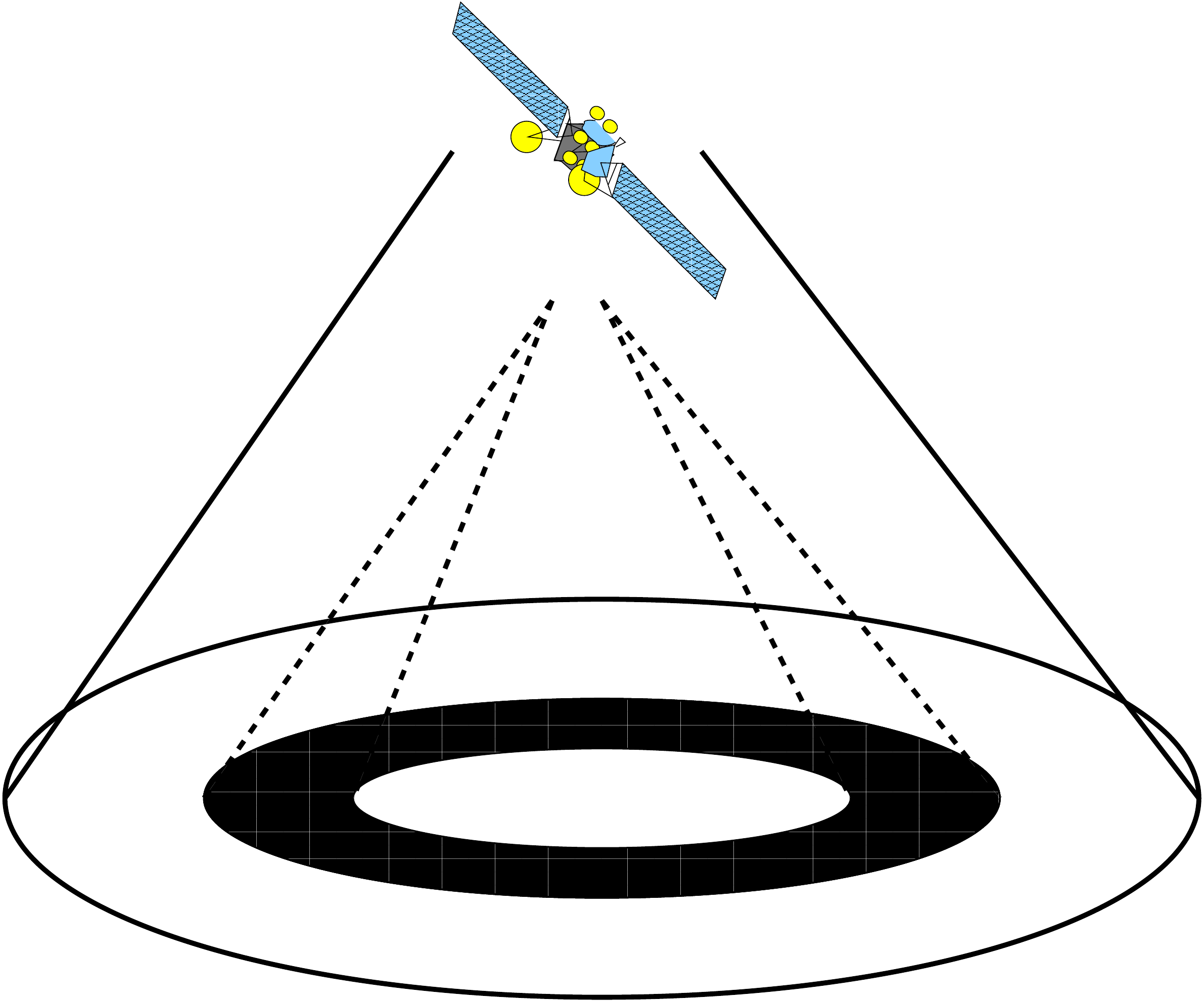}
\caption{Satellite broadcasting area}
\label{spot_sat}
\end{figure}

To evaluate the effective potential of our proposal for real systems, we present a model to estimate the SNR distribution of the receivers for an AWGN channel. For this, we consider the set of receivers located in a given spot beam of a geostationary satellite broadcasting in the Ka band. The model takes into account two main sources of attenuation: the relative location of the terminal with respect to the center of (beam) coverage and the weather. Concerning the attenuation due to the location, the idea is to set the SNR at the center of the spot beam $\text{SNR}_{max}$ and use the radiation pattern of a parabolic antenna to model the attenuation. An approximation of the radiation pattern is 
\begin{equation}
G(\theta) = G_{max} \left( 2\frac{J_1 \left( \sin(\theta) \frac{\pi D}{\lambda} \right)}{\sin(\theta) \frac{\pi D}{\lambda}} \right)^2,
\label{eq_rayonnement}
\end{equation}
where $J_1$ is the first order Bessel function, $D$ is the antenna diameter and $\lambda=c/f$ is the wavelength \cite{antenna}. In our simulations, we use $D=1.5\text{ m}$ and $f=20 \text{ GHz}$. Moreover, we consider a typical multispot system where the edge of each spot beam is 4 dB below the center of coverage. Assuming a uniform repartition of the population, the proportion of the receivers experiencing an attenuation between two given values is the ratio of the ring area over the disk as shown in \figurename~\ref{spot_sat}. The ring area is computed knowing the satellite is geostationary and using (\ref{eq_rayonnement}).

\begin{figure}[!ht]
\centering
\includegraphics[width = 0.9\columnwidth]{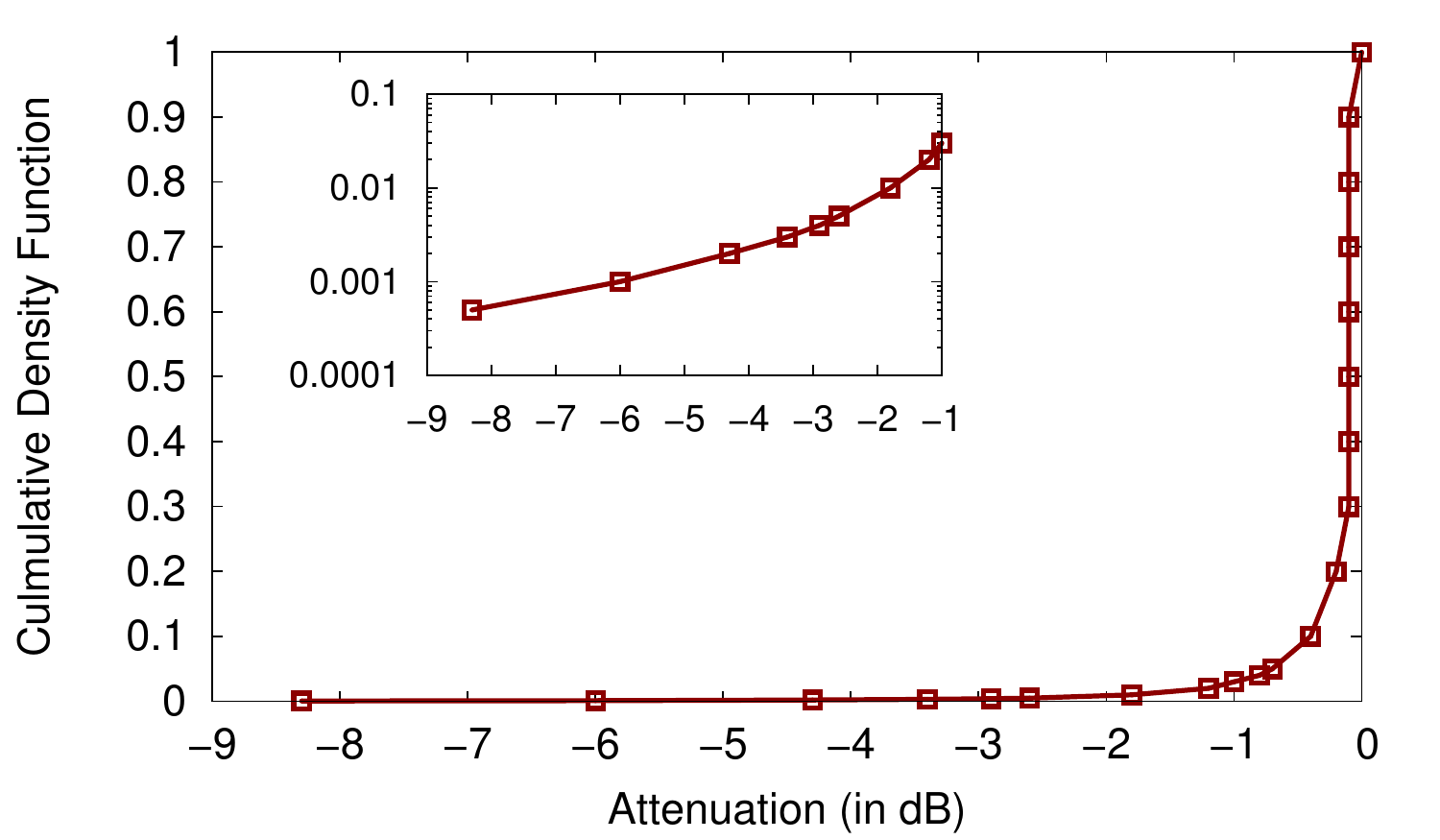}
\caption{Attenuation distribution (due to weather)}
\label{distrib_s2}
\end{figure}

\figurename~\ref{distrib_s2}, provided by the Centre National d'Etudes Spatiales (CNES), presents the attenuation distribution of the Broadcasting Satellite Service (BSS) band. More precisely, it is a temporal distribution for a given location in Toulouse, France. In our work, we assume the SNR distribution for the receivers in the beam coverage at a given time is equivalent to the temporal distribution at a given location.

Finally, our model combines the two attenuations previously described to estimate the SNR distribution. From a set of receivers, we first compute the attenuation due to the location. Then, for each receiver we draw the attenuation caused by the weather according to the previous distribution (see \figurename~\ref{distrib_s2}).

\subsection{Results for the AWGN channel}

\begin{figure}[!ht]
\centering
\includegraphics[width = 0.675\columnwidth]{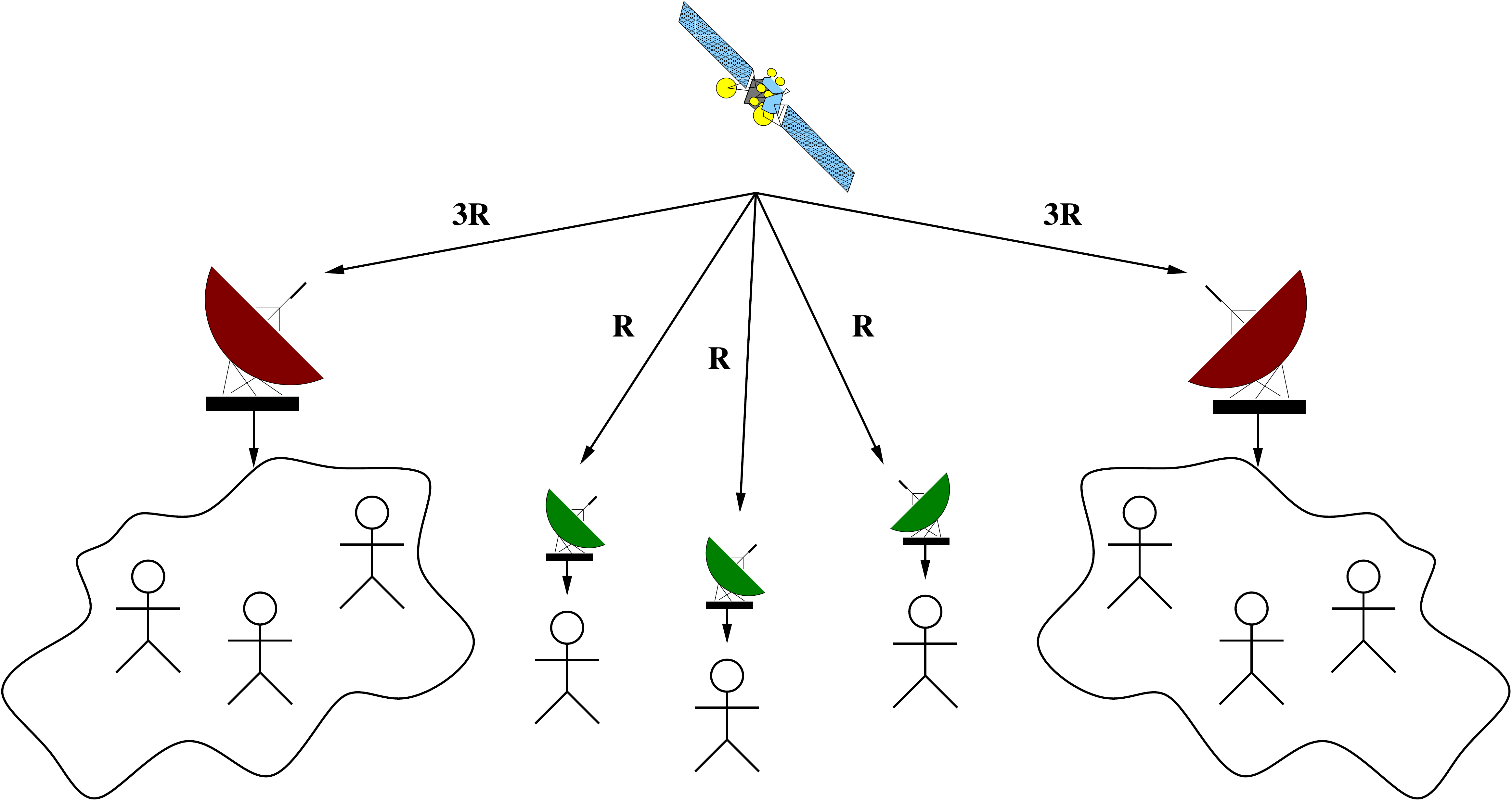}
\caption{Broadcast channel with two kinds of terminals}
\label{two_classes_terminals}
\end{figure}

Two scenarios are considered. In the first one, all the terminals have the same figure of merit $G/T$ (gain to system noise temperature); this scenario is called the homogeneous case. In the second one, we consider two subsets of receivers, one with personal terminals, the other with professional terminals (heterogeneous case). Professional terminals forward the service and not the signal, to some receivers in a local area network. We assume that the professional terminals experience a received SNR 5 dB higher than the personal terminals. They allow an increase in SNR diversity which generally leads to better performance when using the hierarchical modulation time sharing. We consider that only one receiver is served by a personal terminal whereas there are several users behind a professional terminal. In both scenarios, we want to offer the same average rate to all the receivers. Thus the rate dedicated to one professional terminal is proportional to the number of receivers served by this terminal. \figurename~\ref{two_classes_terminals} shows a broadcast channel with two kinds of terminals (large antennas represent professional terminals) where each receiver has an average rate $R$.

\subsubsection{Homogeneous terminals}
\begin{figure}[!ht]
\centering
\includegraphics[width = 0.9\columnwidth]{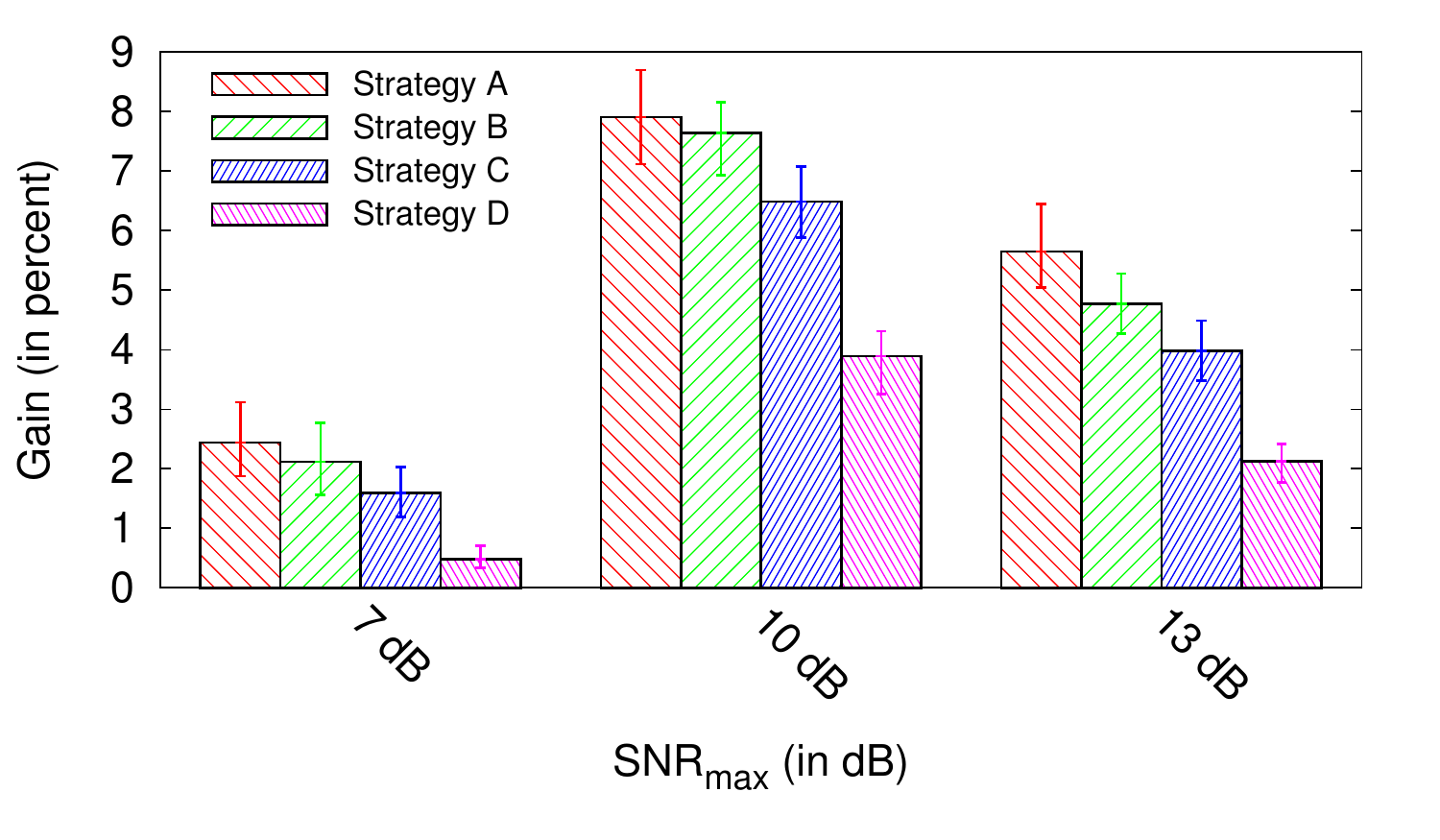}
\caption{Average rate gains for the homogeneous case with 500 receivers}
\label{s2_results}
\end{figure}

\figurename~\ref{s2_results} presents the gains (in terms of average rate) of hierarchical modulation time sharing compared to classical time sharing for a broadcasting area with 500 receivers. For each simulation, the SNR value of each receiver is drawn according to the distribution presented above. Note that this SNR is fixed over all times for a given simulation. We also assume that the transmitter has knowledge of the SNR at the receivers. In practice, this corresponds to a system that implements ACM. For one system configuration (i.e., the parameter $\text{SNR}_{max}$ is set), we present the average, minimum and maximum gains over 100 simulations for the four grouping strategies. 

First of all, hierarchical modulation time sharing outperforms classical time sharing scheme regardless the grouping strategy used. In fact, the hierarchical modulation adds some new operating points and thus can only improve the performance.

For each configuration, strategies A and B give the best results with a slight advantage for strategy A, which obtains a gain of more than 9\%  for $\text{SNR}_{max} = 10$ dB. This is consistent with the results presented in \figurename~\ref{rate_gain}, where the highest gains are obtained when the SNR difference between the two receivers is large. Moreover, strategy D, which minimizes the SNR difference, appears to be the worst scheme. The results also point out that the random strategy performs well. Thus, hierarchical modulation time sharing combined with a clever grouping strategy allows the obtention of intermediate gains between strategies A and D. In addition, strategies A, B and C do not require intensive computation to group the receivers and this can be done in real-time which is interesting for satellite standards. 

\begin{figure}[!ht]
\centering
\includegraphics[width = 0.9\columnwidth]{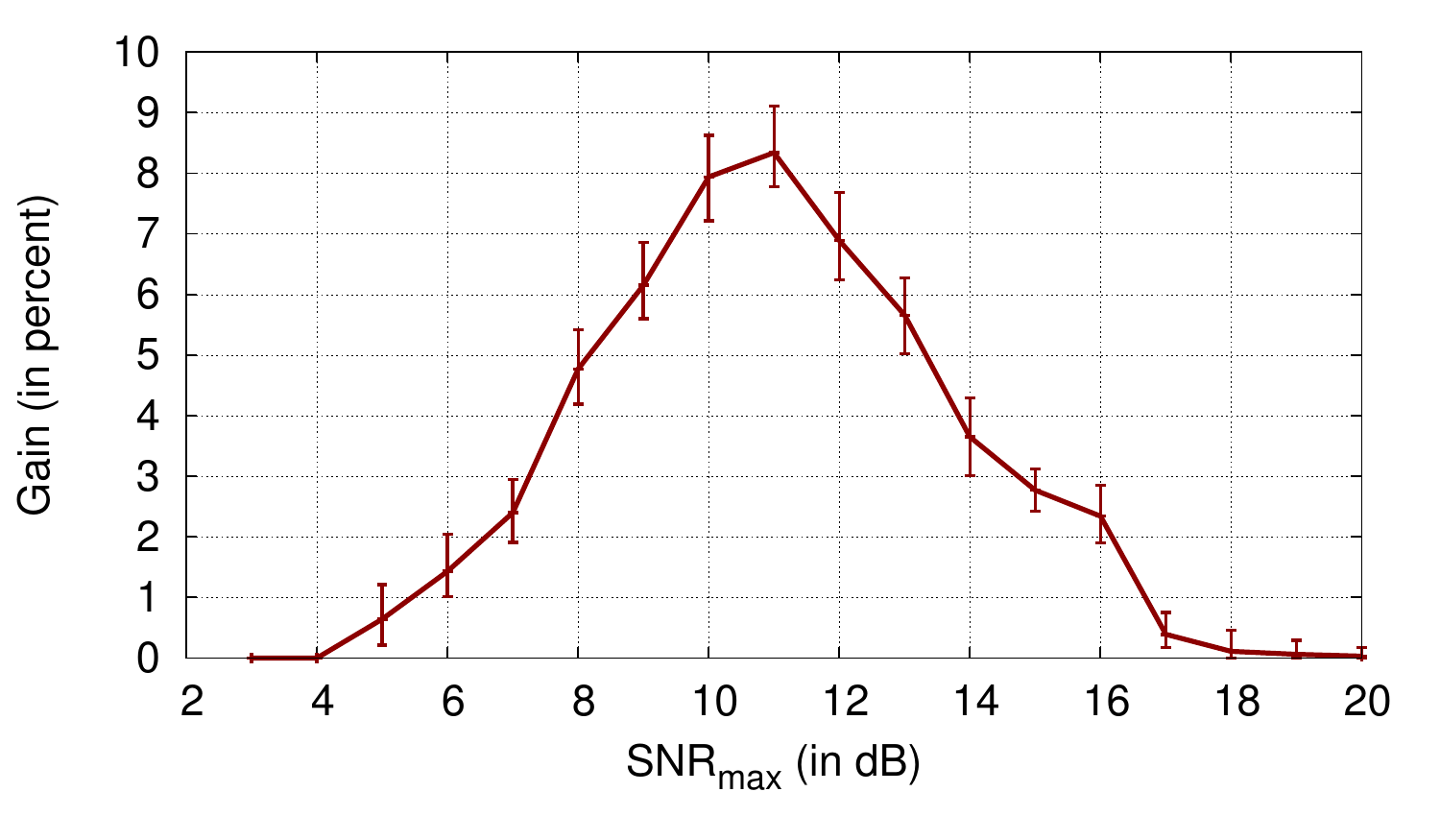}
\caption{Average rate gain vs $\text{SNR}_{max}$ (Strategy A)}
\label{gain_vs_snr}
\end{figure}

The best results are obtained when $\text{SNR}_{max}=10\text{ dB}$. In fact, hierarchical modulation time sharing is useful only in a SNR interval. \figurename~\ref{gain_vs_snr} presents the gains of strategy A according to $\text{SNR}_{max}$ for a large range of $\text{SNR}_{max}$ values. For low SNR values, the LE stream is often not able to decode any coding rate. This explains why there is no gain for low SNR values. An idea to resolve this phenomenon is to allocate more energy to the LE stream, but in that case, the performance of the HE stream deteriorates, too. For large SNR values, classical time sharing uses the largest coding rate possible. For instance, consider two receivers with a SNR greater than 13.13 dB which corresponds to the decoding threshold of data encoded with the LDPC code of rate 9/10 associated with 16-APSK modulation \cite{dvbs2}. The classical time sharing strategy allocates the same fraction of time $t=0.5$ to both receivers. For hierarchical modulation time sharing, one of the receiver decodes the HE stream, and the other the LE stream. In the best case, each stream can decode the code rate 9/10. Each receiver uses the channel all the time but the HE and LE streams only carry two bits. Therefore, hierarchical modulation time sharing can not outperform the classical scheme here. This example illustrates why hierarchical modulation time sharing does not increase the performance for large values of $\text{SNR}_{max}$. A solution would be to use a higher order modulation, for instance a 32-APSK modulation.    

\begin{figure*}[!t]
\centering
\includegraphics[width = 0.75\textwidth]{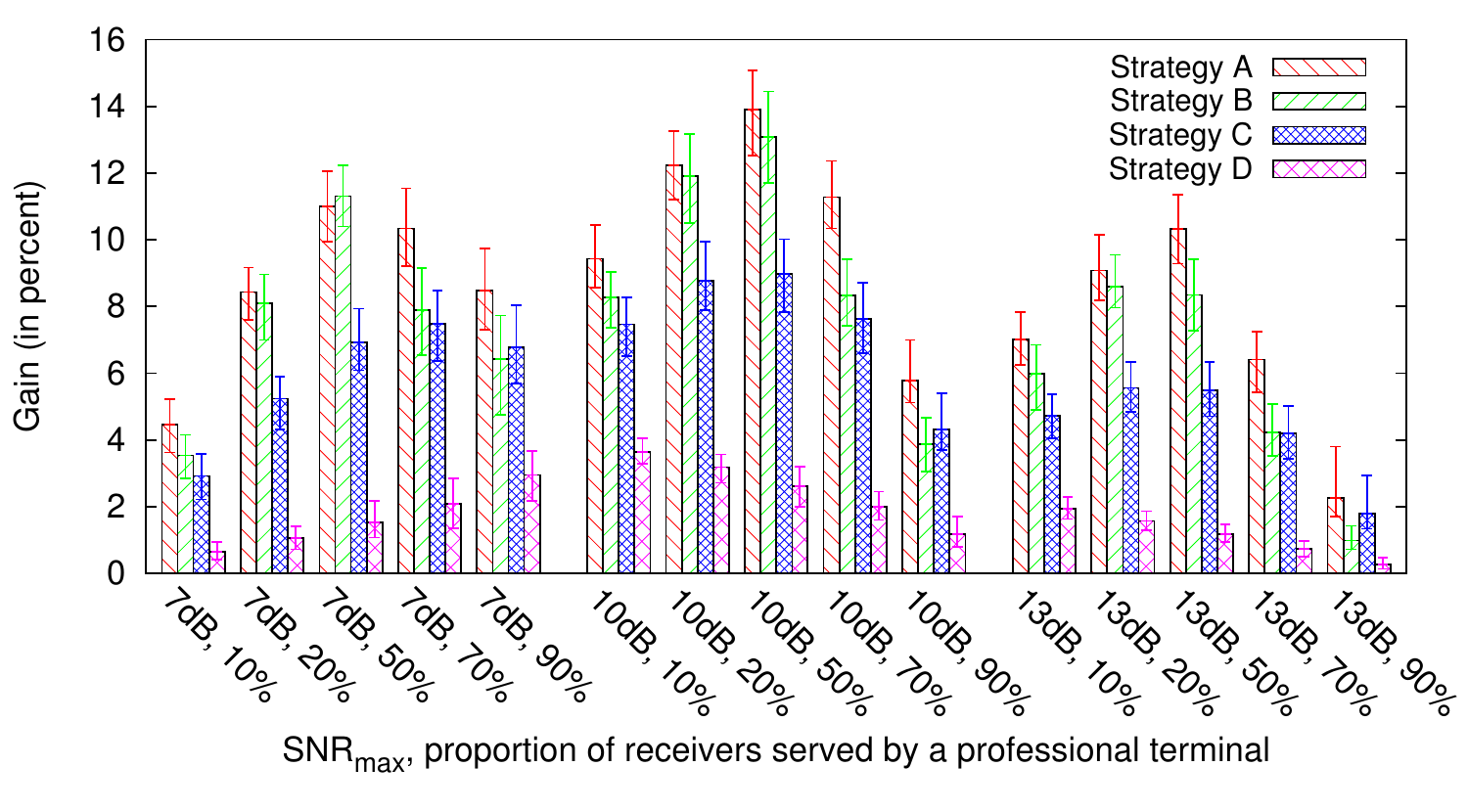}
\caption{Average rate gains for the heterogeneous case with 500 receivers}
\label{s2_results_2}
\end{figure*}

\subsubsection{Heterogeneous terminals}
We investigate the scenario where personal and professional terminals are used simultaneously. As for the homogeneous case, the transmitter has knowledge of the SNR at the receivers. The aim is still to provide the same rate for each receiver. We recall that the professional terminals experience a received SNR 5 dB higher than the personal terminals and the rate dedicated to one professional terminal is proportional to the number of receivers served by this terminal (see \figurename~\ref{two_classes_terminals}).

In this scenario, the performance depends on the proportion of receivers served by a professional terminal and $\text{SNR}_{max}$. Here, $\text{SNR}_{max}$ corresponds to the SNR experienced by personal terminals at the center of the spot beam (with clear sky conditions). \figurename~\ref{s2_results_2} presents the gains according to the proportion of users served by a collective terminal. Here again, the simulations involve 500 receivers and we present the results over 100 simulations. First of all, the gains are mostly better than in \figurename~\ref{s2_results}. This is in accordance with \figurename~\ref{rate_gain} as the presence of professional terminals increases the average SNR difference. Then, for a given $\text{SNR}_{max}$, the maximum gain takes place when 50\% of the receivers are served by a professional terminal, which corresponds to the maximum possible SNR diversity. This result is consistent with the work presented in \cite{icc10}. Finally, compared to \figurename~\ref{s2_results}, the results are worse in two cases, when $\text{SNR}_{max}=10\text{ or } 13\text{ dB}$ and 90\% of the receivers are served by a professional terminal. In these particular cases, we are not really increasing the SNR diversity, but rather the average SNR. Thus, the performance when $\text{SNR}_{max}=10\text{ dB}$ and 90\% of the receivers are served by professional terminals is similar to the performance observed in \figurename~\ref{s2_results_2} when $\text{SNR}_{max}=15\text{ dB}$ (assuming that a professional terminal experiences 5 dB better than a personal terminal).

\section{Conclusion}\label{part5}

In this paper, we use time sharing and hierarchical modulation together to increase the throughput of a broadcast channel. We first propose the hierarchical 16-APSK to generalize the use of hierarchical modulation for the DVB-S2 standard. To the best of our knowledge, the hierarchical 16-APSK has not been extensively studied. Here, we chose the constellation parameters according to an energy argument. Then we presented how to compute achievable rates for our scheme. We introduced several strategies to group the users in pairs. We proposed two scenarios including homogeneous and non-homogeneous terminals and showed that a gain of roughly 15\% can be achieved (in the best case) by a strategy grouping the receivers with the greatest SNR difference.

In this paper, we studied the case where all the receivers obtain the same rate. Future work will extend our work to any rate policy. We also expect to study gains using hierarchical modulation in other standards (e.g., terrestrial standards).

\section*{Acknowledgment}
The authors wish to thank Caroline Amiot-Bazile for sharing the SNR distributions presented in Section~\ref{part4}.

\appendices
\section{Resolution of the energy equation}\label{resolution_equation}

Consider the equation
\begin{equation}
\rho_{he}  = \frac{\left( 1+\gamma (1+2\cos\theta) \right)^2}{4(1+3\gamma^2)},
\label{eq_energie_appendix}
\end{equation}
where $\rho_{he} \geqslant 0.5$ is known, $\gamma \geqslant 1$ and $\theta \geqslant 0$. In order to solve (\ref{eq_energie_appendix}), we transform the equation as follows
\begin{align}
\cos\theta & = \frac{1}{2} \left( \frac{\sqrt{4\rho_{he}(1+3\gamma^2)}-1}{\gamma} -1 \right) \nonumber \\
		   & = f(\gamma, \rho_{he}).
\label{eq_energie_2}
\end{align}

The term $\cos\theta$ is a function that depends on $\gamma$ and $\rho_{he}$. We denote this function $f(\gamma,\rho_{he})$. We now consider when the condition $-1 \leqslant f(\gamma, \rho_{he}) \leqslant 1$ is verified in order to use the $\arccos$ function. The derivative of $f$ shows that the function $f(\gamma, \rho_{he})$ is an increasing function of $\gamma$ when $\rho_{he}$ is set. Using the facts that $\gamma = R_2/R_1 \geqslant 1$ and $\rho_{he} \geqslant 1/2$, we find
\begin{align}
f(\gamma, \rho_{he}) &\geqslant f(1, \rho_{he}) \nonumber \\
					 &= \frac{1}{2} \left( 4\sqrt{\rho_{he}} -2 \right) \nonumber \\
					 &\geqslant \sqrt{2} - 1 
\end{align}
Thus, the inequality $-1 \leqslant f(\gamma, \rho_{he})$ is always true. We now study an upper bound of $f$. First of all, we have the following relation
\begin{equation}
f(\gamma, \rho_{he}) \xrightarrow[\gamma \to +\infty]{} \frac{1}{2} (2\sqrt{3\rho_{he}}-1).
\end{equation}

The right term is an increasing function in $\rho_{he}$ and equals 1 for $\rho_{he}=0.75$. Thus, for all $\rho_{he} \leqslant 0.75$, the condition $-1 \leqslant f(\gamma, \rho_{he}) \leqslant 1$ is true and the $\arccos$ function can be used in (\ref{eq_energie_2}). The solution of (\ref{eq_energie_appendix}) for $\rho_{he} \leqslant 0.75$ is
\begin{equation}
\mathcal{S}_{\rho_{he}} = \left\{ \left( \gamma, \arccos \left( f(\gamma, \rho_{he}) \right) \right) | \gamma \geqslant 1 \right\}.
\label{sol1_appendix}
\end{equation}

When $\rho_{he} > 0.75$, $\gamma$ must stay bounded in order to verify $f(\gamma, \rho_{he}) \leqslant 1$. To determine the limit value $\gamma_{lim}$, we have to solve the equation
\begin{align}
f(\gamma, \rho_{he}) = 1 
&\Leftrightarrow \frac{1}{2} \left(\frac{\sqrt{4\rho_{he}(1+3\gamma^2)}-1}{\gamma} -1 \right) = 1 \nonumber \\
& \Leftrightarrow (12\rho_{he}-9)\gamma^2 -6 \gamma + (4\rho_{he}-1) = 0.
\label{quadratic}
\end{align}

Equation (\ref{quadratic}) is a quadratic equation with discriminant $\Delta = 192\rho_{he}(1-\rho_{he})$. The solutions are
\begin{equation}
s_{1,2} = \frac{6 \pm \sqrt{192\rho_{he}(1-\rho_{he})}}{2(12\rho_{he}-9)}.
\end{equation}

We retain the positive solution,
\begin{equation}
\gamma_{lim} = \frac{3+4\sqrt{3\rho_{he}(1-\rho_{he})}}{3(4\rho_{he}-3)}.
\end{equation}

Finally, the solution of (\ref{eq_energie_appendix}) for $\gamma > 0.75$ is,
\begin{equation}
\mathcal{S}_{\rho_{he}} = \left\{ \left( \gamma, \arccos \left( f(\gamma, \rho_{he}) \right) \right) | 1 \leqslant \gamma \leqslant \gamma_{lim} \right\}.
\label{sol2}
\end{equation}

\begin{figure}[ht!]
\centering
\includegraphics[width=0.8\columnwidth]{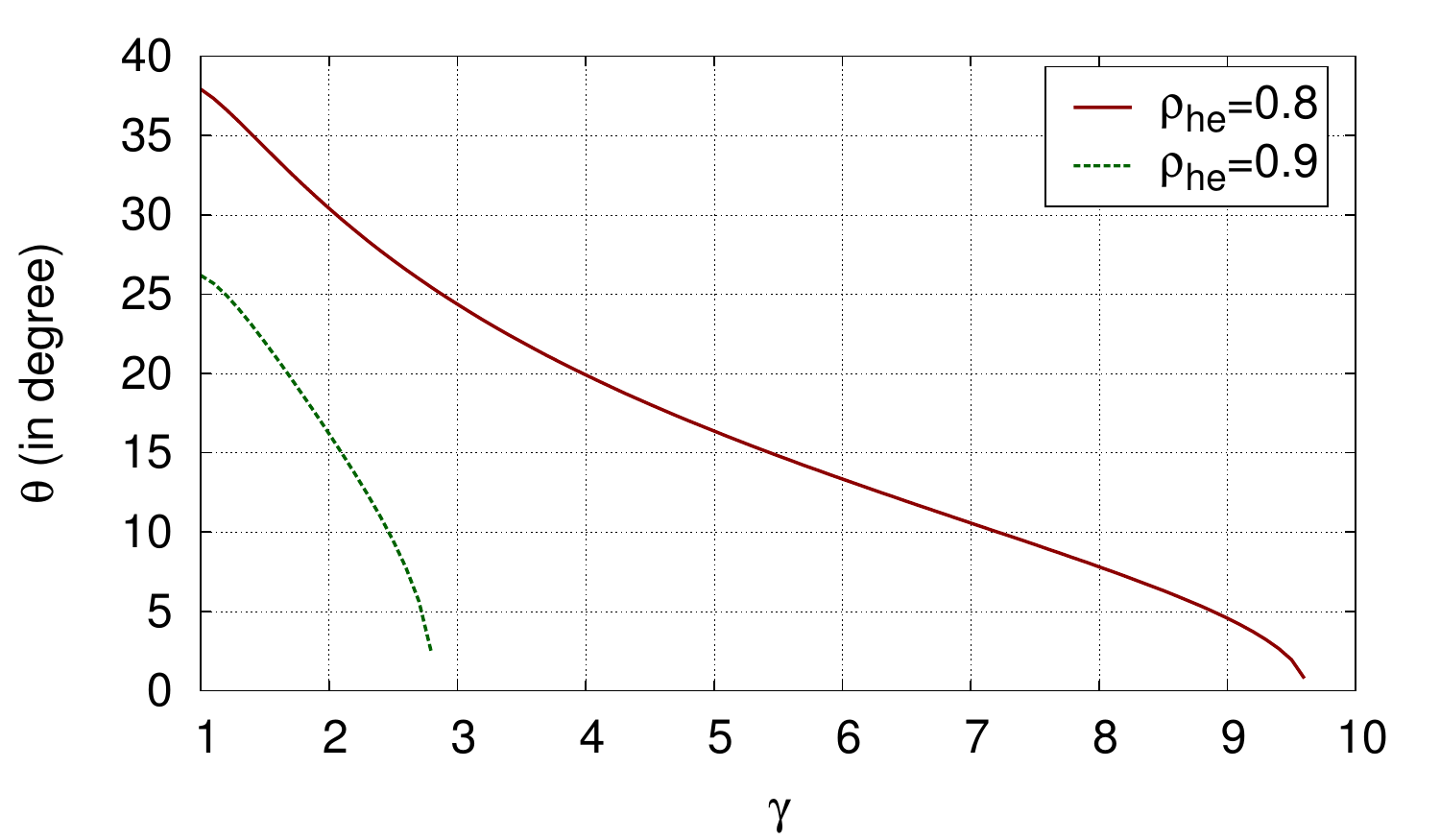}
\caption{Examples of $\mathcal{S}_{\rho_{he}}$}
\label{sol_eq_nrj}
\end{figure}

\figurename~\ref{sol_eq_nrj} presents two examples of $\mathcal{S}_{\rho_{he}}$ with different values of $\rho_{he}$. When $\rho_{he}$ increases, the symbols in one quadrant tend to become closer. For instance, when $\gamma=1$, we find that $\theta=38^\circ$ for $\rho_{he}=0.8$ and $\theta=26^\circ$ for $\rho_{he}=0.9$. Thus, the symbols are closer in the case $\rho_{he}=0.9$. This implies that the HE stream is easier to decode, whereas the LE stream requires a good reception to be decoded.

\section{Hierarchical 16-APSK performance}\label{appendixB}

\begin{figure}[!ht]
\centering
\subfloat[Estimated HE stream decoding thresholds]{\includegraphics[width=0.8\columnwidth]{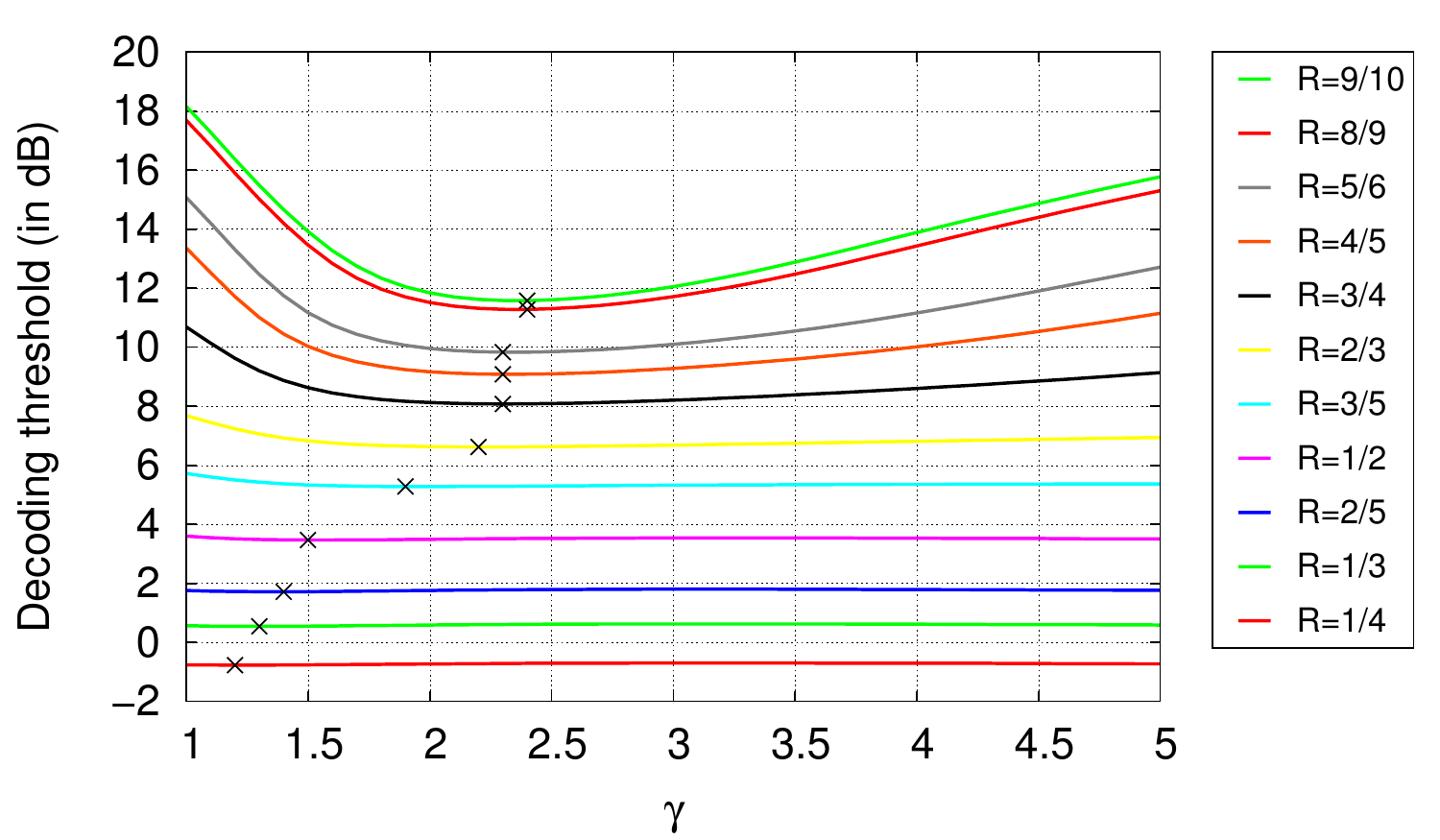}%
\label{estimated_perf_hp_80}}

\subfloat[Estimated LE stream decoding thresholds]{\includegraphics[width=0.8\columnwidth]{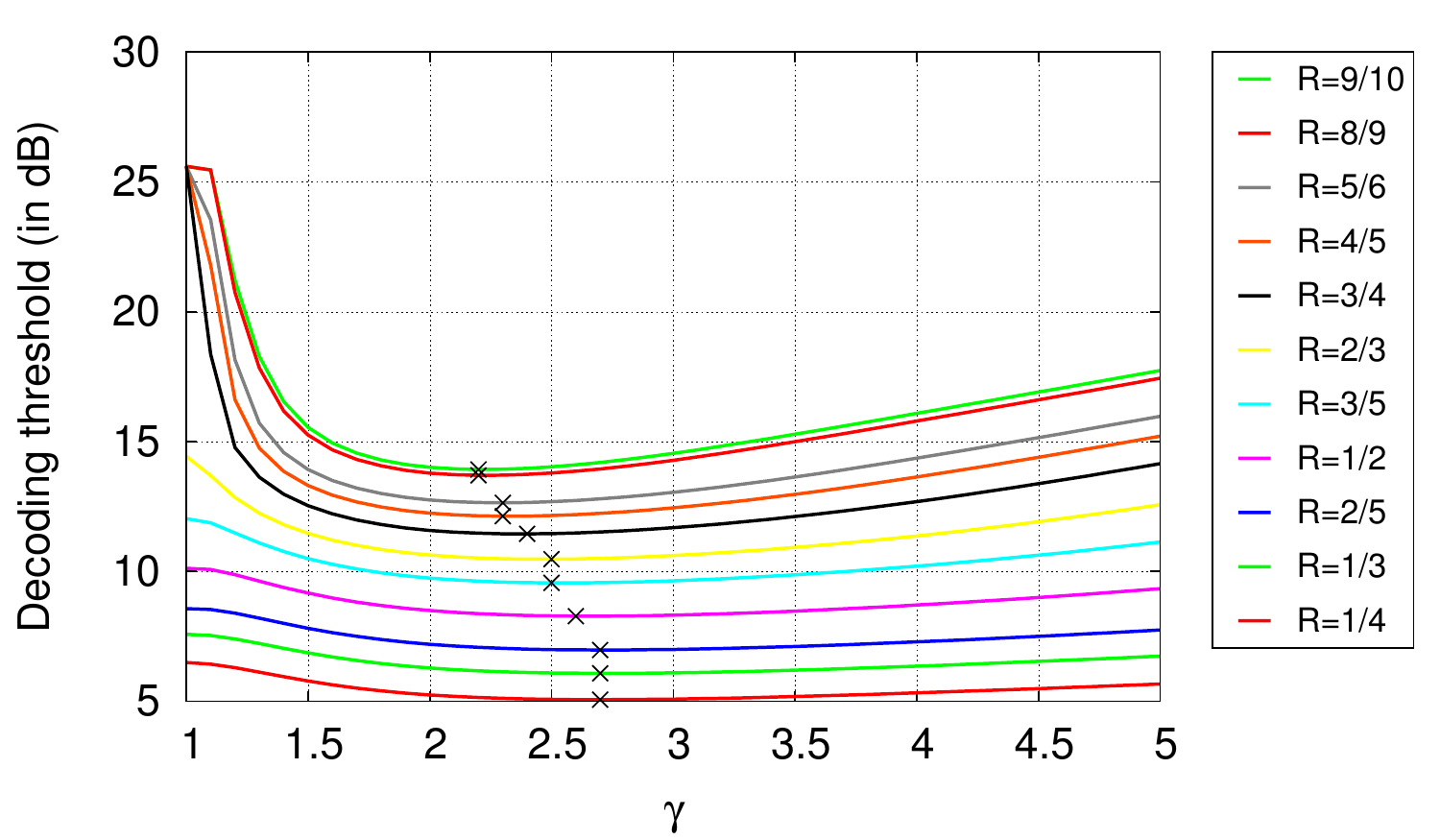}%
\label{estimated_perf_lp_80}}

\caption{Estimated performance of the hierarchical 16-APSK ($\rho_{he}=0.8$)}
\label{estimated_perf}
\end{figure}

We develop in this section the method used to choose the $(\gamma, \theta)$ pair for the hierarchical 16-APSK once the parameter $\rho_{he}$ has been set. We decide to keep only one $(\gamma, \theta)$ pair per $\rho_{he}$ value, as simulations to obtain the performance are time-consuming. For a given $\rho_{he}$, the decoding thresholds for all the coding rates in function of $\gamma$ are estimated using the method described in \cite{wts}. This allows us to obtain a fast approximation of all the decoding thresholds. For instance, \figurename~\ref{estimated_perf} presents the curves obtained for $\rho_{he}=0.8$, where the crosses correspond to the minimum of each curve. Note that the mathematical resolution of (\ref{eq_energie_appendix}) allows large values of $\gamma$ which is not realistic in practical systems. In our work, we decide to upper bound $\gamma$ by $\min (5, \gamma_{lim})$.

Next we chose to adopt the $(\gamma, \theta)$ pair that minimizes the average decoding threshold for the HE stream over all the code rates. \figurename~\ref{estimated_perf} shows that this solution does not significantly penalize the LE stream. With the estimated performance, we pick the $(\gamma, \theta)$ pair according to the previous criteria. Table~\ref{adopted_pairs} present the adopted pairs.
\begin{table}[!ht]
\renewcommand{\arraystretch}{1.1}
\caption{Adopted $(\gamma, \theta)$ values}
\label{adopted_pairs}
\centering
\begin{tabular}{c||c|c|c|c} 
\hline
$\rho_{he}$ & 0.75 & 0.8 & 0.85 & 0.9 \\
\hline
$\gamma$ & 2.8 & 2.3 & 1.9 & 1.6 \\
\hline
$\theta$ & 31.5 & 28.4 & 25.1 & 20.9 \\
\hline 
\end{tabular}
\end{table}

\begin{figure*}[!ht]
\centerline{\subfloat[HE stream performance, $\rho_{he}=0.75$]{\includegraphics[width=0.4\textwidth]{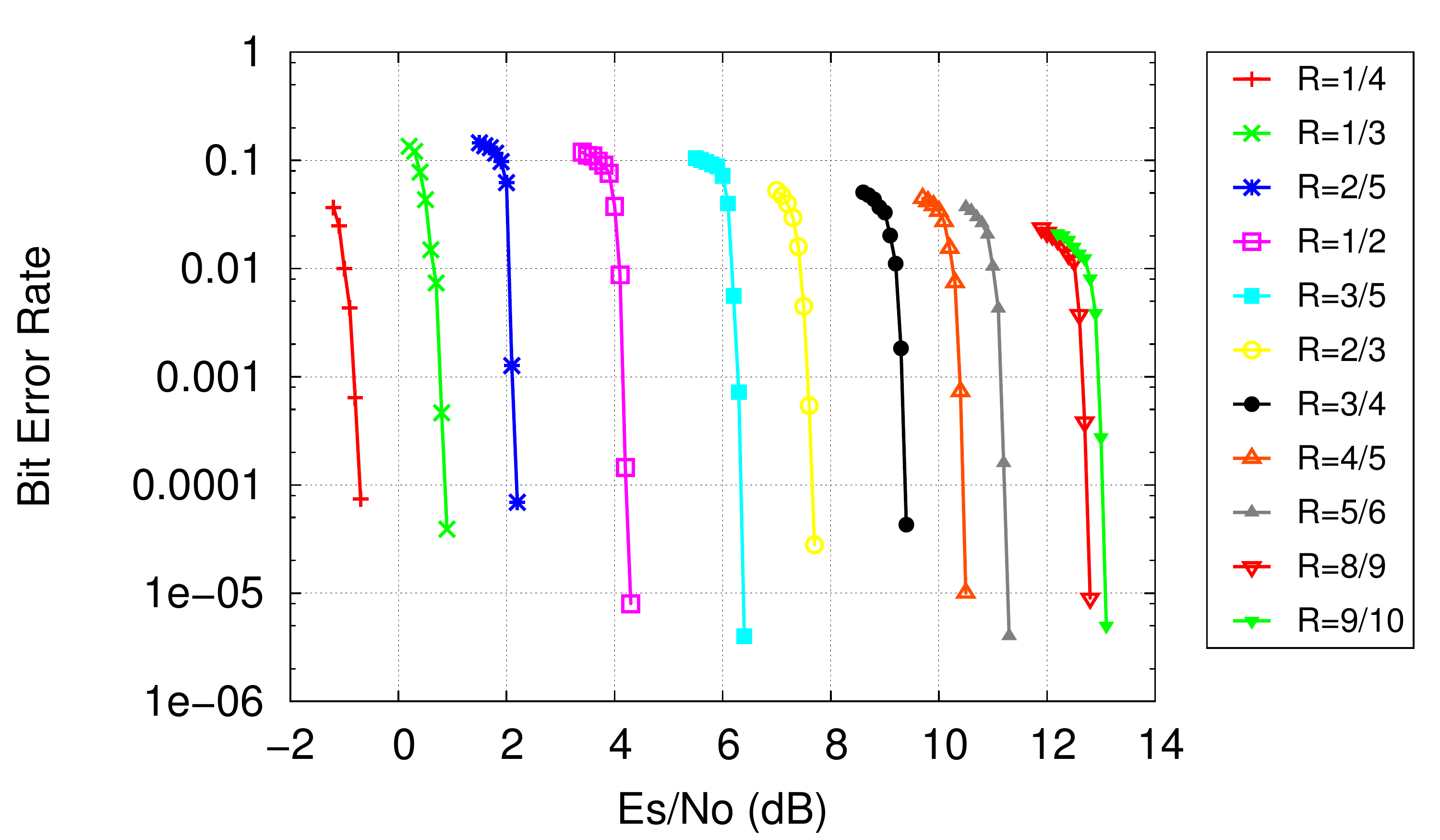}%
\label{perf_hp_75}}%
\hfil
\subfloat[LE stream performance, $\rho_{he}=0.75$]{\includegraphics[width=0.4\textwidth]{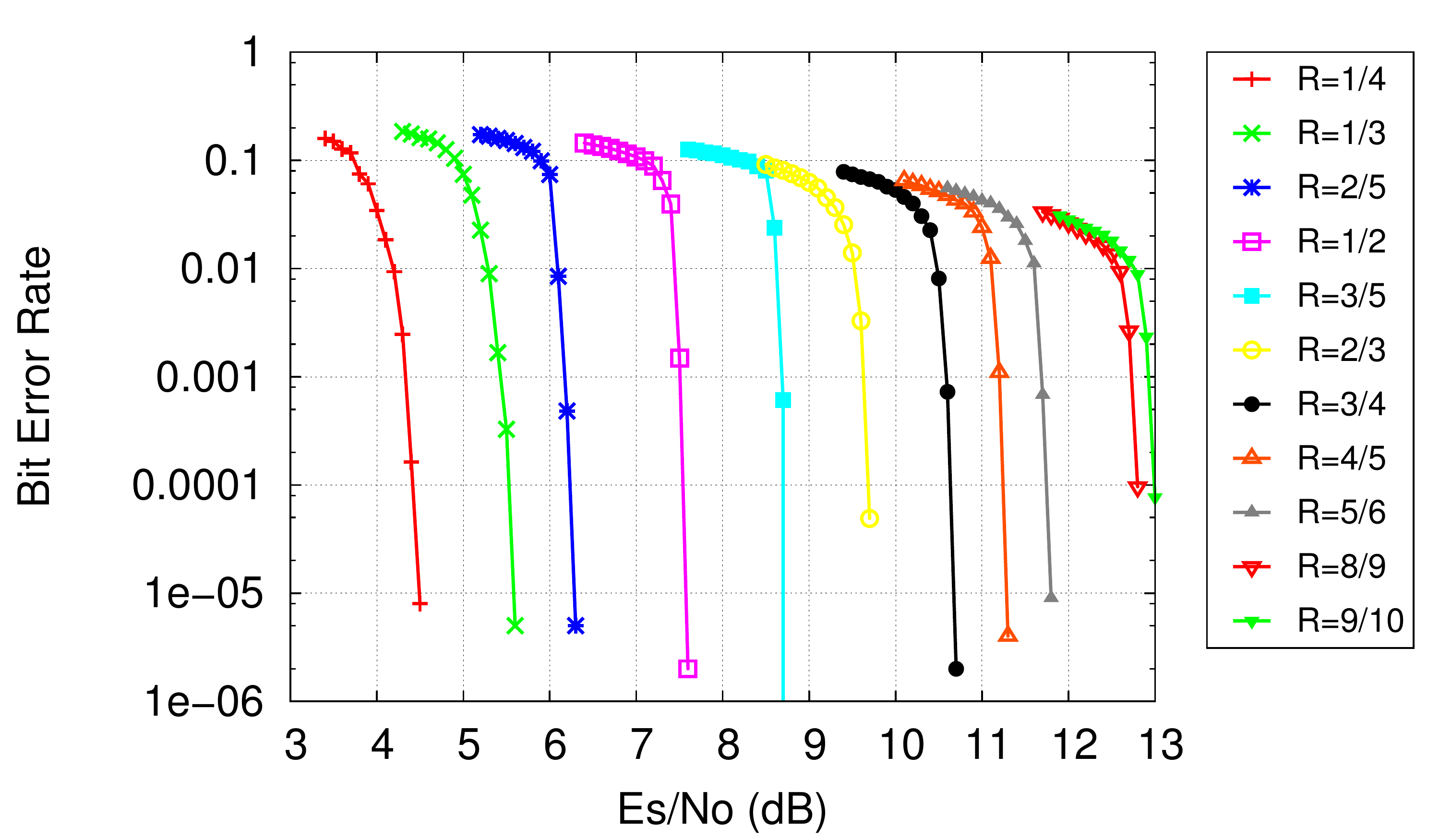}%
\label{perf_lp_75}}}%
\centerline{\subfloat[HE stream performance, $\rho_{he}=0.8$]{\includegraphics[width=0.4\textwidth]{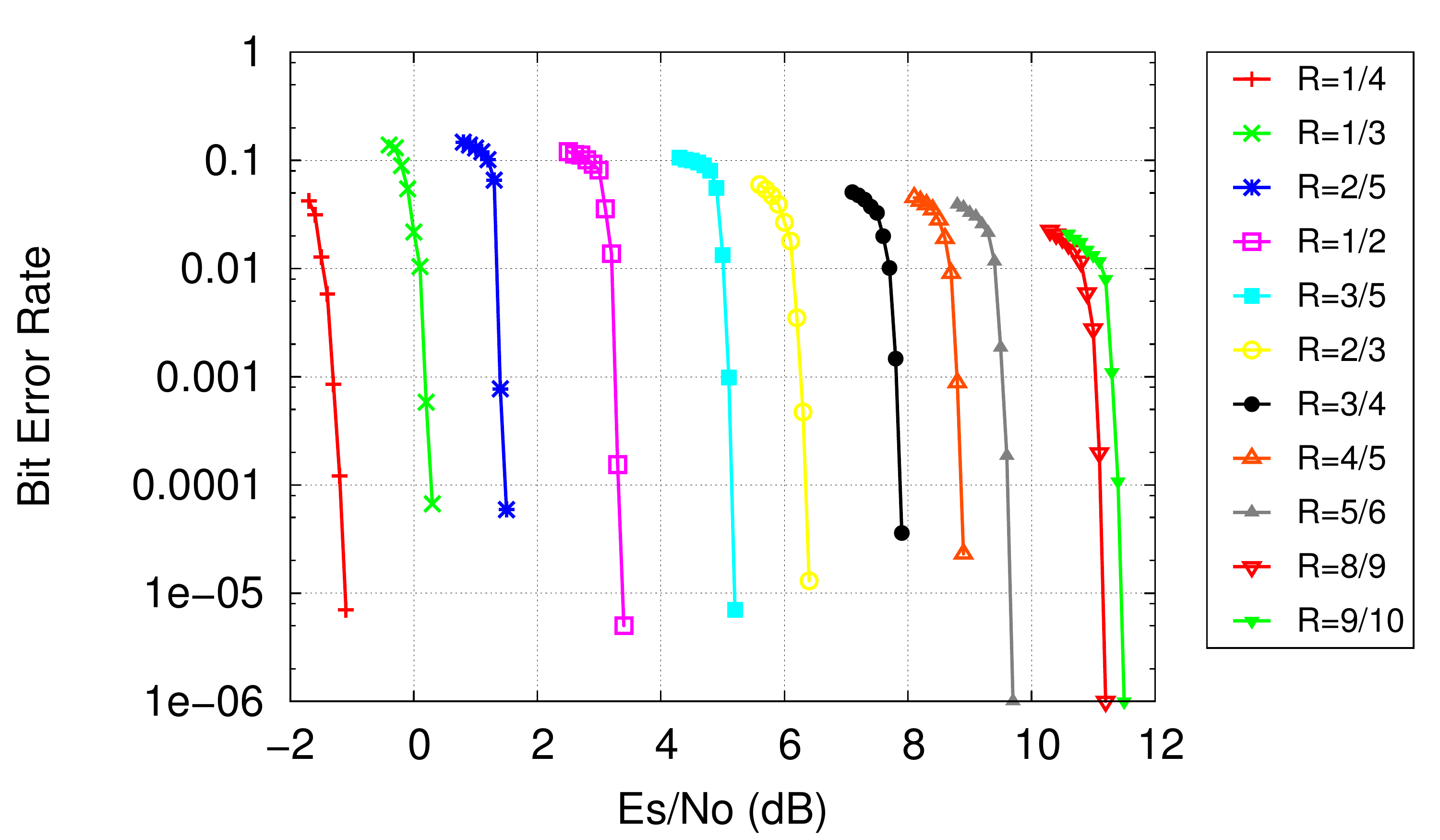}%
\label{perf_hp_80}}%
\hfil
\subfloat[LE stream performance, $\rho_{he}=0.8$]{\includegraphics[width=0.4\textwidth]{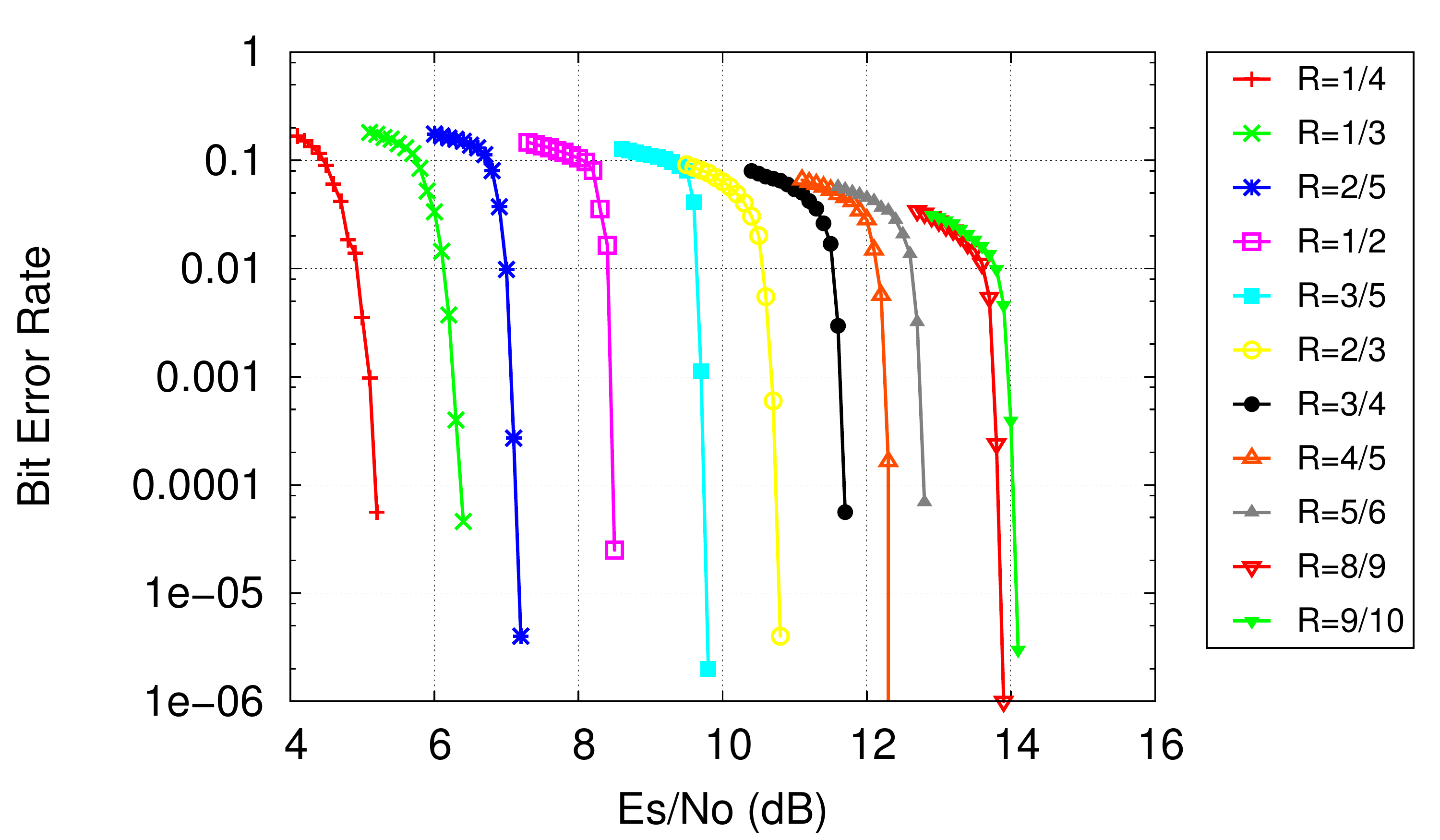}%
\label{perf_lp_80}}}%
\centerline{\subfloat[HE stream performance, $\rho_{he}=0.85$]{\includegraphics[width=0.4\textwidth]{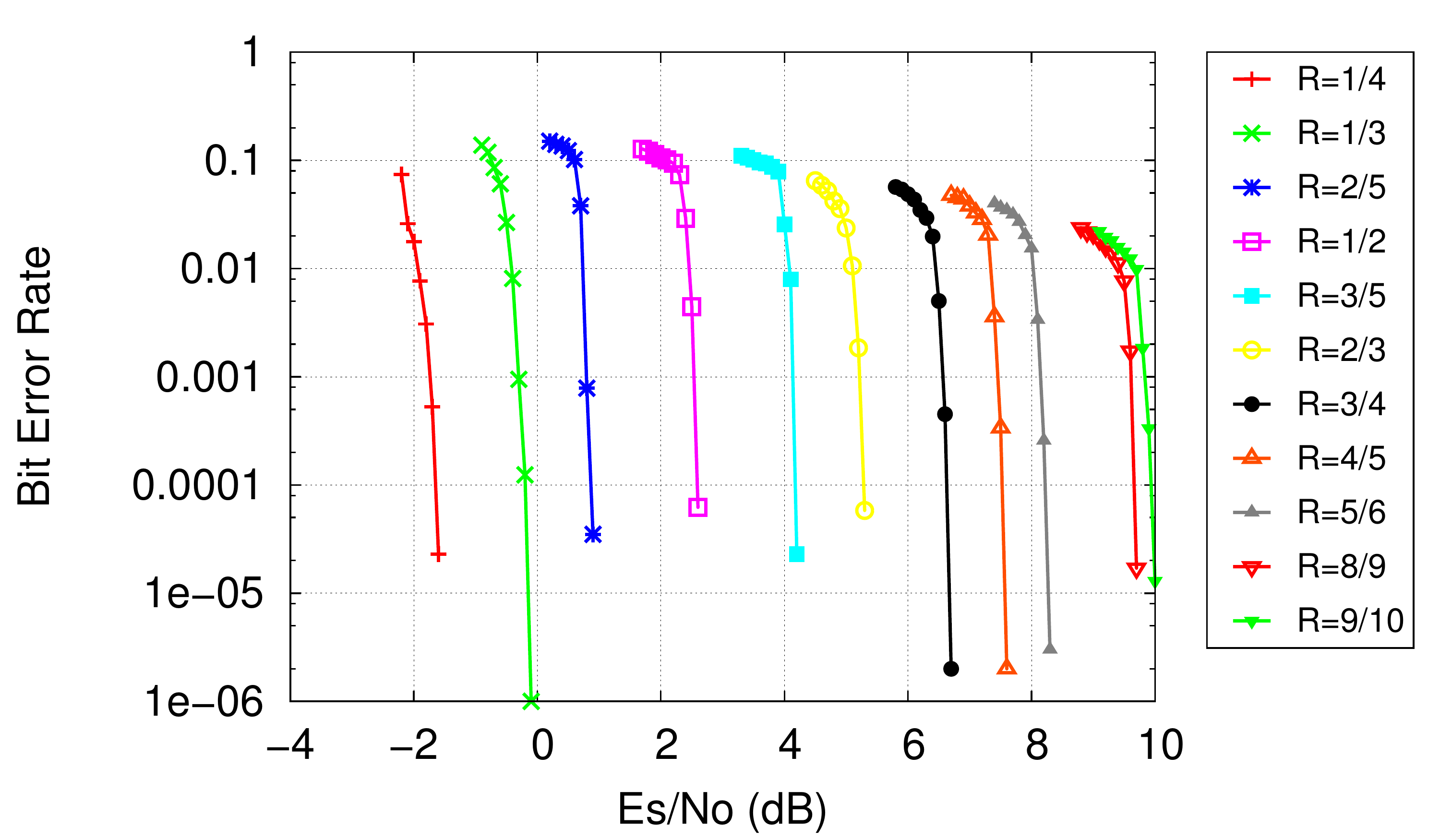}%
\label{perf_hp_85}}%
\hfil
\subfloat[LE stream performance, $\rho_{he}=0.85$]{\includegraphics[width=0.4\textwidth]{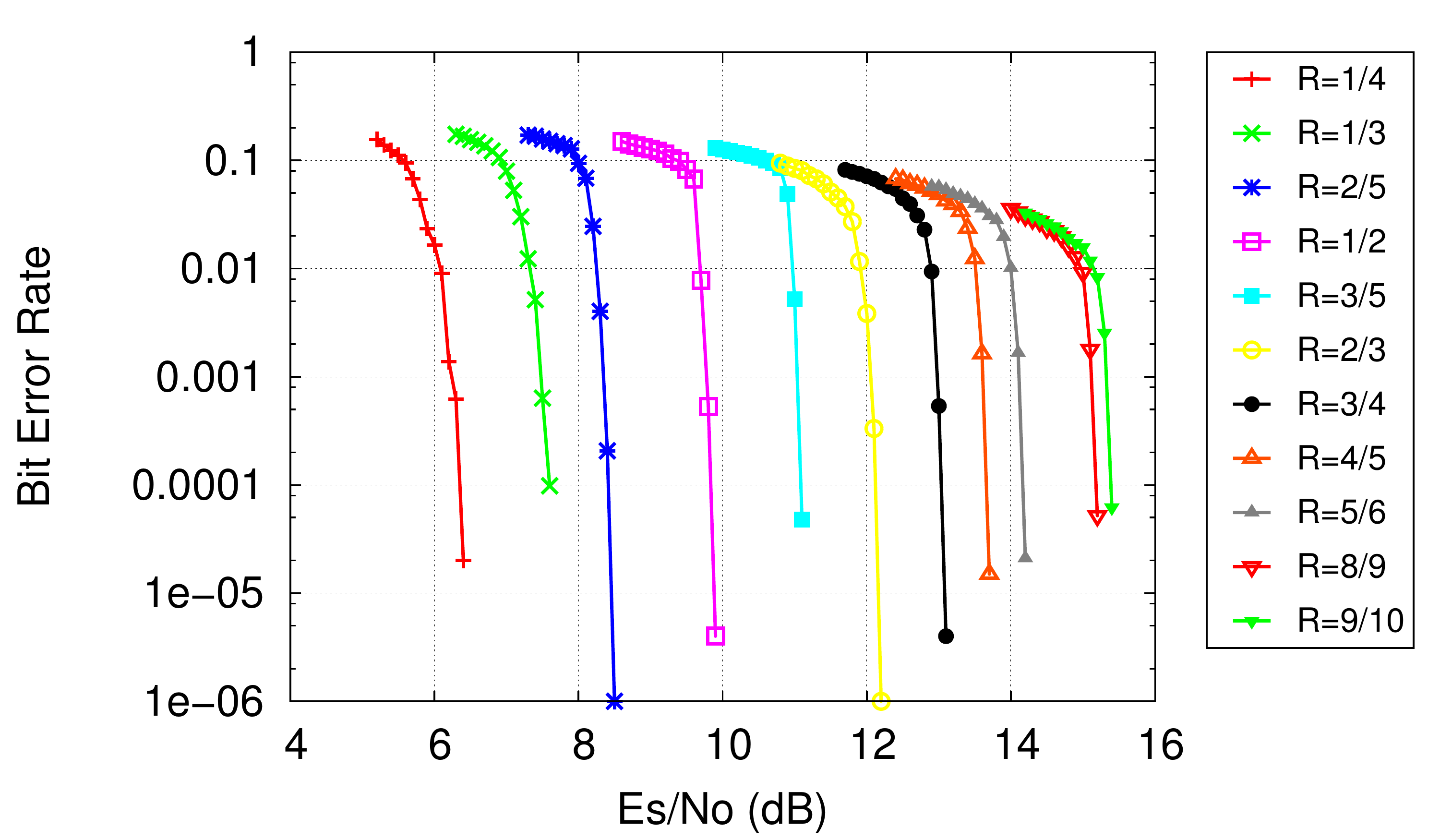}%
\label{perf_lp_85}}}%
\centerline{\subfloat[HE stream performance, $\rho_{he}=0.9$]{\includegraphics[width=0.4\textwidth]{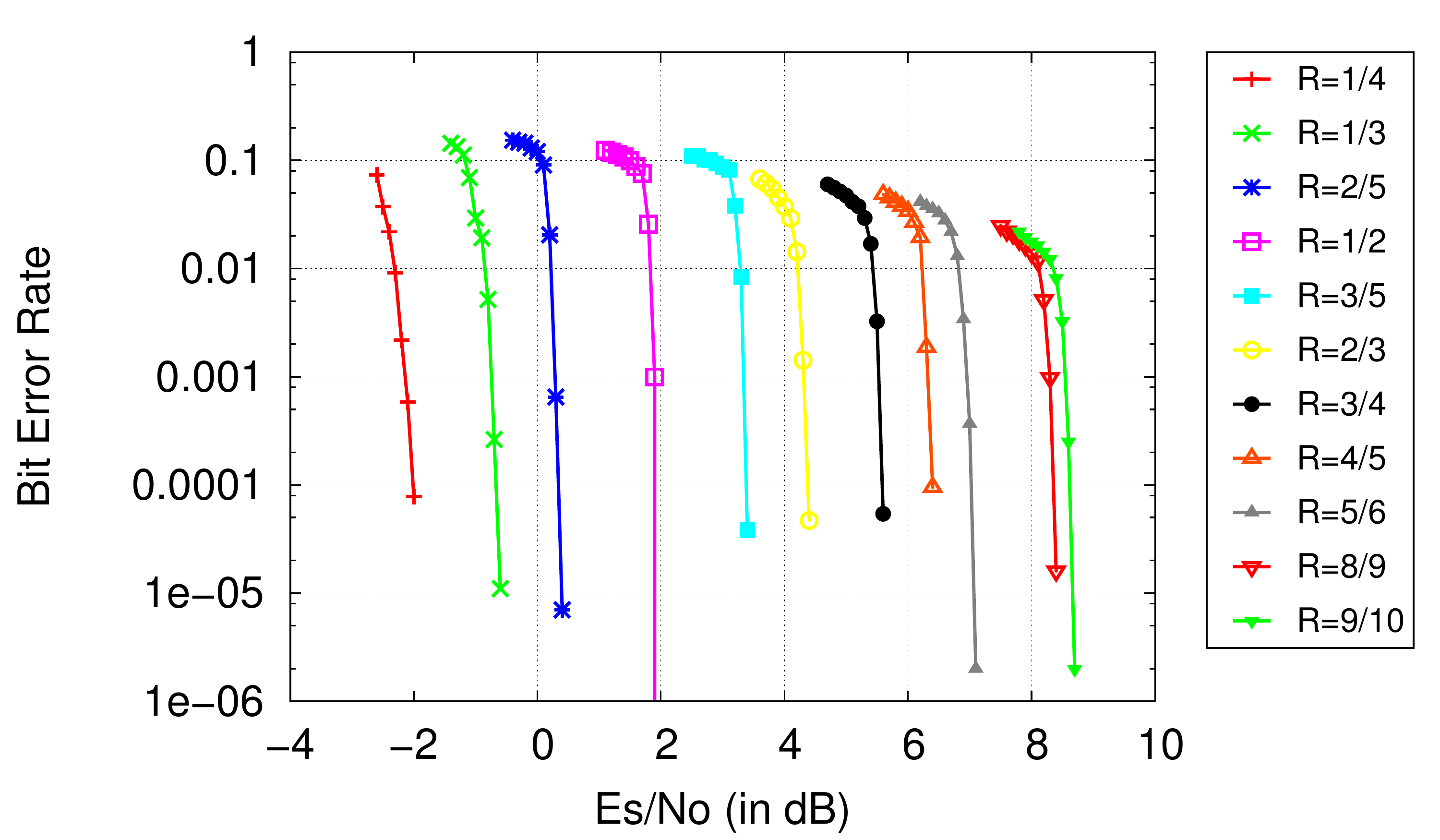}%
\label{perf_hp_90}}%
\hfil
\subfloat[LE stream performance, $\rho_{he}=0.9$]{\includegraphics[width=0.4\textwidth]{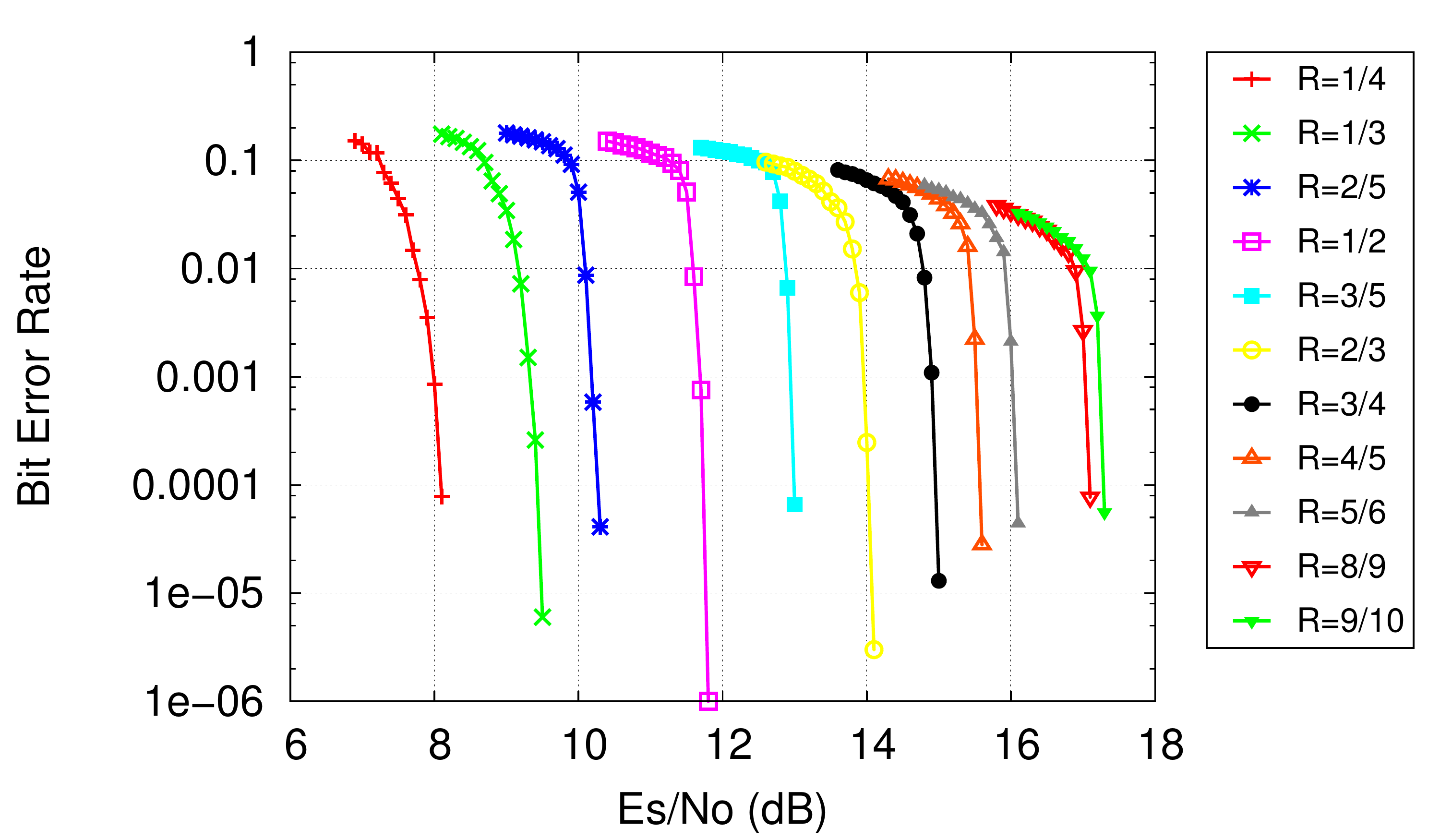}%
\label{perf_lp_90}}}%

\caption{Performance of the hierarchical 16-APSK on an AWGN channel}
\label{perf}
\end{figure*}

Finally, the performance is evaluated with simulations using the Coded Modulation Library \cite{cml} that already implements the DVB-S2 LDPC codes. The LDPC codewords are 64 800 bits long (normal FEC frame) and the iterative decoding stops after 50 iterations if no valid codeword has been decoded. Moreover, in our simulations, we wait after 10 decoding failures before computing the Bit Error Rate (BER). If the BER is less than $10^{-4}$, then we stop the simulation. Our stopping criterion is less restrictive than in \cite{s2} (i.e, a packet error rate of $10^{-7}$) because simulations are time-consuming. However, our simulations are sufficient to detect the waterfall region of the LDPC and then the code performance. \figurename~\ref{perf} presents the BER curves for the HE and LE streams of the hierarchical 16-APSK on an AWGN channel.

\section{Proof of Theorem~\ref{our_theorem}}\label{proof_theorem}

First, we compute an upper bound for the average SNR difference. In (\ref{average_delta}), for all $k$, we have (assuming $i_k \leqslant j_k$)
\begin{equation}
\Delta_{i_k,j_k} = \sum_{m=i_k}^{j_k-1} \Delta_{m, m+1}.
\end{equation}

Thus, $\Delta$ can be expressed in the following form
\begin{equation}
\Delta = \frac{1}{N} \sum_{i=1}^{m-1} a_i \Delta_{i,i+1},
\label{delta_max}
\end{equation}
where $a_i \in \mathbb{N}$ for all $i$. We now try to bound $a_i$. The term $\Delta_{i,i+1}$ in (\ref{delta_max}) only appears when we group a user with a SNR less than or equal to $\text{SNR}_i$ and a user with a SNR greater than or equal to $\text{SNR}_{i+1}$. There are exactly $\sum_{k=1}^{i} n_k$ receivers with $\text{SNR} \leqslant \text{SNR}_i$ and $\sum_{k=i+1}^{m} n_k$ receivers with $\text{SNR} \geqslant \text{SNR}_{i+1}$, so $a_i$ is bounded by
\begin{equation}
a_i \leqslant \min ( \sum_{k=1}^{i} n_k,  \sum_{k=i+1}^{m} n_k).
\label{bound}
\end{equation}

We now prove the proposed scheme reaches this bound, i.e., $a_i = \min ( \sum_{k=1}^{i} n_k,  \sum_{k=i+1}^{m} n_k)$. Let $L$ be the greatest integer such as $\sum_{i=1}^{L}n_i \leqslant N$. The strategy ensures that all the receivers with a SNR less than or equal to $\text{SNR}_{L+1}$ are grouped with a receiver whose SNR is greater than or equal to $\text{SNR}_{L+1}$. Thus, in the computation of the average SNR difference, we verify that
\begin{itemize}
\item $\Delta_{1,2}$ appears $n_1 = \min ( \sum_{k=1}^{1} n_k,  \sum_{k=2}^{m} n_k)$ times.
\item $\Delta_{2,3}$ appears $n_1+n_2 = \min ( \sum_{k=1}^{2} n_k,  \sum_{k=3}^{m} n_k)$ times.
\item ...
\item $\Delta_{L,L+1}$ appears $n_1+n_2+...+n_L = \min ( \sum_{k=1}^{L} n_k,  \sum_{k=L+1}^{m} n_k)$ times.
\end{itemize}
The equality also holds in (\ref{bound}) for the terms $\Delta_{i,i+1}$ with $i \geqslant L+1$. Thus, our strategy allows us to reach the previous bound, which is in fact the maximum average SNR difference.

\ifCLASSOPTIONcaptionsoff
  \newpage
\fi

\nocite{*}
\bibliographystyle{IEEEtran}
\bibliography{biblio}

\end{document}